\pdfoutput=1
\documentclass[preprint,12pt,nopreprintline]{elsarticle}

\usepackage{amsmath,amssymb}
\usepackage{graphicx}
\usepackage{placeins}
\usepackage{booktabs}
\usepackage{array}
\usepackage{xcolor}
\usepackage{listings}
\usepackage{url}
\usepackage{tikz}
\usetikzlibrary{arrows.meta,decorations.markings,calc}
\usepackage[colorlinks=true,linkcolor=blue!50!black,citecolor=blue!50!black,
            urlcolor=blue!50!black]{hyperref}

\newtheorem{proposition}{Proposition}
\newdefinition{remark}{Remark}
\newproof{proof}{Proof}

\newcommand{\R}{\mathbb{R}}
\newcommand{\bx}{\mathbf{x}}
\newcommand{\by}{\mathbf{y}}
\newcommand{\bz}{\mathbf{z}}
\newcommand{\bv}{\mathbf{v}}
\newcommand{\bu}{\mathbf{u}}
\newcommand{\bc}{\mathbf{c}}
\newcommand{\br}{\mathbf{r}}
\newcommand{\bs}{\mathbf{s}}
\newcommand{\bn}{\mathbf{n}}
\newcommand{\be}{\mathbf{e}}
\newcommand{\bA}{\mathbf{A}}
\newcommand{\bQ}{\mathbf{Q}}
\newcommand{\bI}{\mathbf{I}}
\newcommand{\bJ}{\mathbf{J}}
\newcommand{\bS}{\mathbf{S}}
\newcommand{\bxi}{\boldsymbol{\xi}}
\newcommand{\bomega}{\boldsymbol{\omega}}
\newcommand{\boeta}{\boldsymbol{\eta}}
\newcommand{\bvh}{\hat{\mathbf{v}}}
\newcommand{\Phih}{\hat{\Phi}}
\newcommand{\Op}{\Omega^{+}}
\newcommand{\dd}{\mathrm{d}}
\DeclareMathOperator{\SOgroup}{SO}
\DeclareMathOperator{\Id}{Id}

\journal{Journal of Computational Physics}

\begin{document}

\begin{frontmatter}

\title{Cyclic changes of frame for designing non-retracing verification cases for numerical interface tracking}

\author[tuda]{Dieter Bothe\corref{cor1}}
\ead{bothe@mma.tu-darmstadt.de}
\author[tuda]{Tomislav Mari\'{c}\corref{cor1}}
\ead{maric@mma.tu-darmstadt.de}
\cortext[cor1]{Corresponding authors.}

\affiliation[tuda]{organization={Mathematical Modeling and Analysis,
                                 Technical University of Darmstadt},
                   country={Germany}}

\begin{abstract}
A forward--backward deformation test supplies an exact final interface, but
retracing can allow errors from the two halves to cancel. We construct
alternative tests by superposing a cyclic change of frame on the same base
deformation. The frame returns to its starting pose, preserving the exact
final phase and the deformation history up to rigid placement. A small
returning translation can keep trajectories close to the retraced ones.
A full rotation about a fixed centre or axis gives closed paths with
nonzero winding for material labels that stay off that centre or axis.
For incompressible base fields the transformed velocity is divergence-free. We construct discretely
solenoidal face fluxes for geometric volume-of-fluid advection by
circulating the transformed potential. In three TwoPhaseFlow benchmarks,
with a shared time grid for each pair, a full-turn frame increased the
final cell-average shape discrepancy in all 30 comparisons, with a median
factor of $1.50$.
The fitted convergence exponent decreased in each of the six
benchmark--reconstruction comparisons over the tested grids.
\end{abstract}


\end{frontmatter}

\section{Introduction}

\label{sec:intro}

Deformation tests challenge a numerical interface with thin sheets or
filaments and use a known final phase to measure the resulting error.
Let $\Op(0)$ be the initial phase and $\Phi_0^t$ the flow map of a
prescribed velocity $\bv$ on the region traversed by the phase. If
\begin{equation}
  \Phi_0^T(\bx_0)=\bx_0 \quad \text{for all relevant material points } \bx_0 ,
  \label{eq:identity}
\end{equation}
then $\Op(T)=\Op(0)$ exactly. The initial geometry supplies the final reference
without a separate reference advection calculation. Rigid rotation of a
slotted disc \cite{zalesak1979} is one familiar return test. In widely used
deformation tests \cite{leveque1996,enright2002,rider1998}, a forward flow
stretches the phase and a reversed flow brings it home. In that case every
material label is in the same place at $t$ and $T-t$. Numerical errors from
the two halves can then oppose one another at the endpoint.

A cyclic change of frame lets us choose how the phase returns. At every time
$t$, apply a rigid motion $E_t$ to the whole configuration produced by the
base flow at that time. The stationary mesh sees a different placement of
the same deforming shape. If $E_0=E_T=\Id$, every material point still returns
exactly. A small returning translation changes the path by an arbitrarily
small amount. Full revolutions, nonuniform angular speeds and combined
rotations and translations give different paths and time histories,
subject to confinement of the swept phase.

There are precedents in tracer transport. Nair and Lauritzen
\cite[Section~3.2.3]{nair2010} combined a reversing deformation on the sphere
with a synchronized zonal revolution. Chen et al.\
\cite[Section~3.3]{chen2017} used translation through a periodic Cartesian
domain, where the frame closes modulo the domain length. We formulate the
construction for returning rigid motions in Euclidean space, characterize
mirrored-time separation of material labels, and give a discretely
solenoidal realization for geometric volume-of-fluid (VOF) advection.

\begin{figure}[p]
\centering
\includegraphics[width=\textwidth]{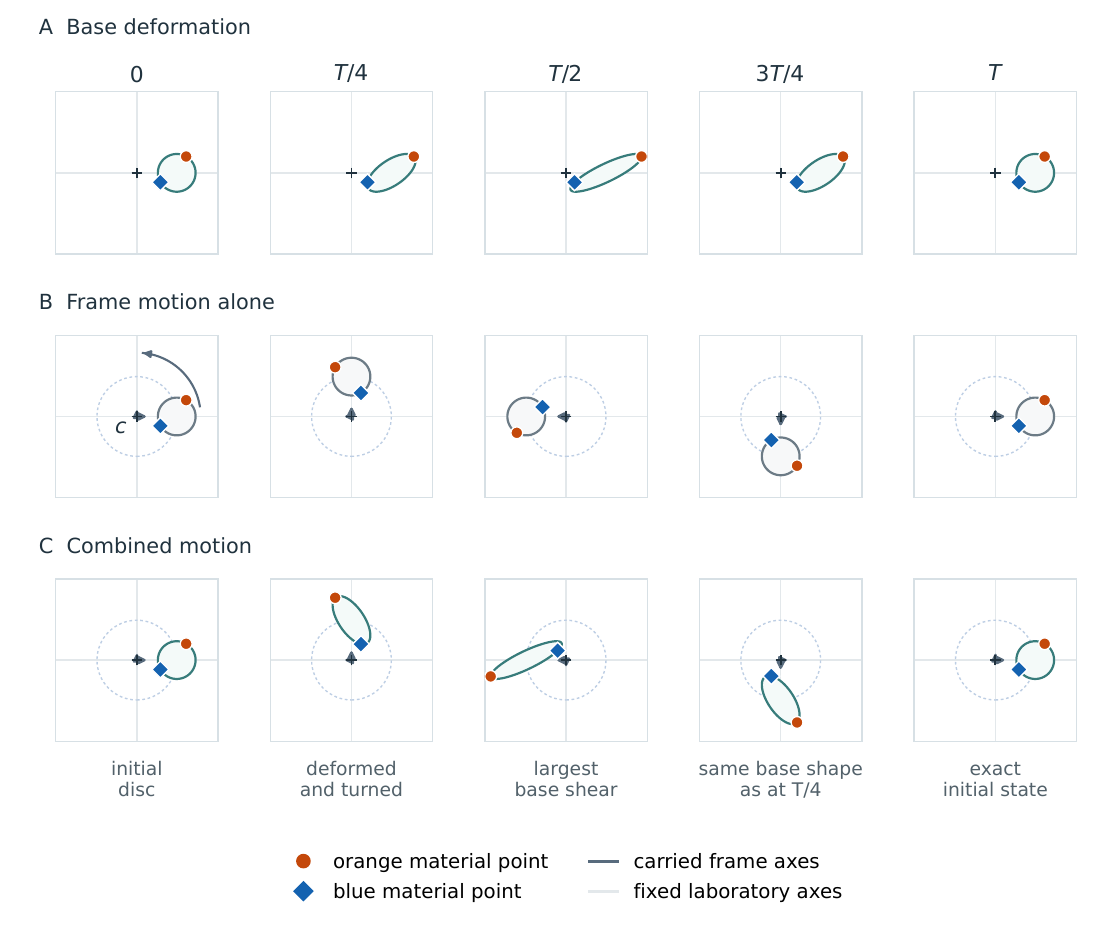}
\caption{Two returning motions and their combination. Read each row from
left to right. \textbf{A:} the base flow deforms a disc and then undoes
that deformation. \textbf{B:} the rigid frame alone carries the undeformed
disc through one full turn. \textbf{C:} at each time, apply the frame pose
in B to the deformed shape in A. Orange circles and blue diamonds identify
the two material points throughout. Slate-grey arrows are the carried
reference axes. The light grey axes remain fixed in the laboratory. The dotted circle marks the
disc-centre orbit under the frame motion alone and serves as a placement
guide in C. At $T/4$ and $3T/4$, each coloured point occupies the same
position in A and appears on opposite sides of the rotation centre in C.
At $T$, every row recovers its initial configuration. The snapshots follow
directly from the analytic example in \eqref{eq:illustration}.}
\label{fig:points}
\end{figure}

Figure~\ref{fig:points} separates the two motions before combining them.
The base flow changes the shape; the rigid frame changes its placement.
Compare the three entries in any time column: the bottom shape has the
same deformation as the top shape, carried by the frame pose in the
middle row. The coloured points make this correspondence visible.
The frame is superposed continuously while the base flow acts. It is not
applied only after the deformation has finished, and the computational
mesh remains stationary.

Section~\ref{sec:cancel1d} shows how midpoint quadrature can give zero
terminal error despite an erroneous intermediate position.
Section~\ref{sec:method} derives the cyclic-frame construction and its
conditions for return and separation. For geometric VOF, we obtain
discretely solenoidal face fluxes by circulating the potential of the
prescribed velocity (Section~\ref{sec:vof}). The paired studies in
Sections~\ref{sec:results}--\ref{sec:leveque} use isoAdvector with plicRDF
and isoAlpha reconstructions in \texttt{TwoPhaseFlow} \cite{scheufler2023}.
Section~\ref{sec:agnostic} summarizes the procedure for other implementations.

\FloatBarrier
\section{Error cancellation in one dimension}
\label{sec:cancel1d}

Consider a point with initial position $x(0)=0$ and prescribed velocity
\begin{equation}
  v(t) = U\cos\!\left(\frac{\pi t}{T}\right),
  \label{eq:v1d}
\end{equation}
Section~\ref{sec:results} uses this temporal profile in the deformation
tests. The velocity reverses at $T/2$, with $v(T-t) = -v(t)$. The exact
trajectory $x(t) = (UT/\pi)\sin(\pi t/T)$ satisfies $x(T-t) = x(t)$ and
returns to $x(T) = 0$.

The midpoint rule
\begin{equation}
  x_{n+1} = x_n + h\,v\bigl(t_n + \tfrac{h}{2}\bigr), \qquad
  t_n = nh, \qquad h = T/N, \qquad N \text{ even},
  \label{eq:midpoint1d}
\end{equation}
reduces to second-order midpoint quadrature for this spatially uniform
velocity. Summing the first $N/2$ steps gives the error at $T/2$,
\begin{equation}
  x_{N/2} - x(T/2) = \frac{U h}{2\sin\bigl(\pi h/(2T)\bigr)} - \frac{UT}{\pi}
                   = \frac{\pi U h^2}{24\,T} + O(h^4),
  \label{eq:half1d}
\end{equation}
At $T$, the midpoints of steps $n$ and
$N-1-n$ lie symmetrically about $T/2$, $t_{N-1-n} + h/2 = T - (t_n + h/2)$, where
the velocity takes opposite values. The error at $T$ therefore vanishes pair by
pair,
\begin{equation}
  x_N - x(T) = h\sum_{n=0}^{N/2-1}
     \Bigl[v\bigl(t_n + \tfrac{h}{2}\bigr) + v\bigl(T - t_n - \tfrac{h}{2}\bigr)\Bigr]
   = 0 ,
  \label{eq:full1d}
\end{equation}
for every even $N$. The terminal error therefore misses the nonzero midpoint
error in \eqref{eq:half1d}.

In one dimension the Euclidean frame of Section~\ref{sec:step2} reduces to
a translation. Consider the returning displacement
\begin{equation}
  z(t) = \beta\,\frac{t^2(T-t)}{T^2}, \qquad z(0) = z(T) = 0 .
  \label{eq:z1d}
\end{equation}
The point now moves with the velocity $v + \dot z$, and its exact position
$x(t) = (UT/\pi)\sin(\pi t/T) + z(t)$ still satisfies $x(T) = 0$, while
$x(T-t)\neq x(t)$ for $\beta\neq0$ and $0<t<T/2$. The translation breaks
coincidence at mirrored times even on a line, where a nonconstant returning
trajectory necessarily revisits positions. At $T$ the terms of $v$ still cancel
as in \eqref{eq:full1d}. The midpoint error comes from the added quadratic
velocity $\dot z$,
\begin{equation}
  x_N - x(T) = h\sum_{n=0}^{N-1}\dot z\bigl(t_n + \tfrac{h}{2}\bigr)
               - \bigl(z(T) - z(0)\bigr)
             = -\frac{h^2}{24}\bigl(\ddot z(T) - \ddot z(0)\bigr)
             = \frac{\beta h^2}{4T} .
  \label{eq:frame1d}
\end{equation}
For $\beta\neq0$ the modified test has a nonzero second-order terminal
error. It comes entirely from the additional velocity, with the original
cosine contribution still cancelling. The expressions \eqref{eq:half1d}--\eqref{eq:frame1d}
are checked with SymPy \cite{meurer2017} by the script
\path{cancellation_1d.py} in the data archive \cite{bothe2026figshare}.

The endpoint error also depends on the choice of frame.
For $z(t)=\varepsilon\sin(2\pi t/T)$ with $\varepsilon\ne0$, the mirrored
positions differ at every interior pair. The midpoint samples of $\dot z$
sum to zero on any uniform grid with at least two steps, since their
phases are equally spaced over one period. On the even grids used above,
the cosine contribution from \eqref{eq:v1d} also vanishes, giving zero
terminal error despite the loss of mirrored-time symmetry.

\section{Construction of a returning advection field}
\label{sec:method}

\subsection{Flow maps}
\label{sec:flowmaps}

For a prescribed velocity $\bv(t,\bx)$, a material trajectory satisfies
\begin{equation}
  \frac{\dd \bx}{\dd t} = \bv\bigl(t, \bx(t)\bigr), \qquad \bx(s) = \bx_s .
  \label{eq:ode}
\end{equation}
The flow map $\Phi_s^t(\bx_s)$ gives the position at time $t$ of a
trajectory starting at $\bx_s$ at time $s$ \cite{maric2020},
\begin{equation}
  \Phi_s^t(\bx_s) = \bx_s + \int_s^t \bv\bigl(\tau, \Phi_s^\tau(\bx_s)\bigr)\,\dd\tau .
  \label{eq:flowmap}
\end{equation}
Applied to every point of a set $V$, it transports the set,
$\Phi_s^t(V) = \{\Phi_s^t(\bx) : \bx \in V\}$, and the phase at time $t$ is
$\Op(t) = \Phi_0^t\bigl(\Op(0)\bigr)$ (Fig.~\ref{fig:points}). Following a
trajectory from $s$ to $r$ and then from $r$ to $t$ is the same as following it
from $s$ to $t$,
\begin{equation}
  \Phi_r^t\bigl(\Phi_s^r(\bx)\bigr) = \Phi_s^t(\bx) .
  \label{eq:compose}
\end{equation}
In particular, integrating \eqref{eq:ode} backwards from $t$ to $s$ returns every
point whenever the required trajectories exist and are unique. Thus $\Phi_t^s$
is the inverse of $\Phi_s^t$ on the corresponding domain
\cite{maric2020}. The benchmarks constructed here have the stronger property
that forward integration over $[0,T]$ gives the identity on the admissible
initial positions. For the final phase reference, it would suffice for the
map to return $\Op(0)$ as a set.

\subsection{Superposing a cyclic change of frame}
\label{sec:idea}

We prescribe the motion between fixed endpoints by composing two motions
that each return at $T$. First take $E_0=\mathrm{Id}$.

The base field $\bvh$ applies the first-half velocity in reverse temporal
order and with opposite sign during the second half. This reverses the
first-half flow. Section~\ref{sec:step1} shows that its flow map $\Phih_0^t$,
with $a := T/2$, satisfies
\begin{equation}
  \Phih_0^T(\bx_0) = \bx_0, \qquad
  \Phih_0^{a+\tau}(\bx_0) = \Phih_0^{a-\tau}(\bx_0), \quad 0\leq\tau\leq a .
  \label{eq:baseidea}
\end{equation}
The equality at mirrored times $a\pm\tau$ is the symmetry that allows
the cancellation in Section~\ref{sec:cancel1d}.

The second motion $E_t$ is a rotation, possibly combined with a translation.
It preserves distances and has the identity pose at both endpoints,
\begin{equation}
  E_0(\bx) = E_T(\bx) = \bx \qquad \text{for every } \bx .
  \label{eq:rigididea}
\end{equation}
An example is the rotation about a point $\bc$ by the angle
$\theta(t) = 2\pi t/T$,
$E_t(\bx) = \bc + \bQ\bigl(\theta(t)\bigr)(\bx - \bc)$ with the rotation matrix
$\bQ(\theta)$: during $[0,T]$ every off-centre point turns once about $\bc$ and is
back at its initial position at $T$ (row B of Fig.~\ref{fig:points}).

At each $t$, apply the frame to the base position $\Phih_0^t(\bx_0)$.
The pose $E_t$ then gives the laboratory trajectory (row C of Fig.~\ref{fig:points})
\begin{equation}
  \bx(t) = \Phi_0^t(\bx_0) = E_t\bigl(\Phih_0^t(\bx_0)\bigr)
  \label{eq:idea}
\end{equation}
Inserting \eqref{eq:baseidea} and \eqref{eq:rigididea}
at $t = T$ gives the exact return of every point,
\begin{equation}
  \Phi_0^T(\bx_0) = E_T\bigl(\Phih_0^T(\bx_0)\bigr) = E_T(\bx_0) = \bx_0 ,
\end{equation}
and therefore of the phase, $\Op(T) = \Phi_0^T\bigl(\Op(0)\bigr) = \Op(0)$.
At every intermediate time the laboratory phase is
$E_t(\widehat{\Omega}^+(t))$, where
$\widehat{\Omega}^+(t)=\Phih_0^t(\Op(0))$ is the base phase. The exact
deformation history is preserved up to rigid motion, including the thickness
of fine structures and the phase topology. The orientation and position
relative to the stationary mesh generally change.

At mirrored times $a-\tau$ and $a+\tau$, the base flow gives the same position,
\begin{equation*}
  \by := \Phih_0^{a-\tau}(\bx_0) = \Phih_0^{a+\tau}(\bx_0),
\end{equation*}
and the laboratory separation is determined by the two frame poses,
\begin{equation}
  \bx(a+\tau) - \bx(a-\tau) = E_{a+\tau}(\by) - E_{a-\tau}(\by) .
  \label{eq:mirroridea}
\end{equation}
For the rotation about $\bc$, the angles at the two times differ by
$\varphi = \theta(a+\tau) - \theta(a-\tau) = 4\pi\tau/T$. With
$\bQ(\theta+\varphi) = \bQ(\theta)\bQ(\varphi)$ and $\br := \by - \bc$,
\eqref{eq:mirroridea} becomes
\begin{equation*}
  \bx(a+\tau) - \bx(a-\tau)
  = \bQ\bigl(\theta(a+\tau)\bigr)\br - \bQ\bigl(\theta(a-\tau)\bigr)\br
  = \bQ\bigl(\theta(a-\tau)\bigr)\bigl(\bQ(\varphi) - \bI\bigr)\br .
\end{equation*}
The rotation $\bQ(\theta(a-\tau))$ does not change lengths. In the plane,
$\br\cdot\bQ(\varphi)\br = \lVert\br\rVert^2\cos\varphi$, so
\begin{equation*}
  \bigl\lVert\bigl(\bQ(\varphi) - \bI\bigr)\br\bigr\rVert^2
  = 2\lVert\br\rVert^2 - 2\,\br\cdot\bQ(\varphi)\br
  = 2\lVert\br\rVert^2(1-\cos\varphi)
  = 4\lVert\br\rVert^2\sin^2(\varphi/2) ,
\end{equation*}
the squared length of the chord between $\br$ and its rotation by $\varphi$.
With $\varphi = 4\pi\tau/T$,
\begin{equation}
  \bigl\lVert \bx(a+\tau) - \bx(a-\tau) \bigr\rVert
  = 2\,\bigl\lvert\sin(2\pi\tau/T)\bigr\rvert\,\lVert \by - \bc\rVert ,
  \label{eq:gapidea}
\end{equation}
which is positive for $0<\tau<a$ if $\by\ne\bc$. At $T/4$ and $3T/4$ the
same base point is rotated by a quarter and three quarters of a turn,
respectively, so these two positions lie opposite one another about $\bc$
(row C of Fig.~\ref{fig:points}). For a planar full turn, every label that avoids the
centre is separated at every interior mirrored pair. Separation at even one
pair suffices to break exact mirrored-time retracing for that label. This
criterion concerns matched times and makes no claim about path intersections
at other times.

A full turn also rules out complete retracing with a different time
parametrization. For a base trajectory that avoids the rotation centre,
choose a continuous polar angle about that centre. The first and second
halves give opposite angle increments. The frame adds $2\pi$, so the
laboratory trajectory has winding number one. An out-and-back traversal
at any speed has winding number zero. In three dimensions the same
argument uses the projection perpendicular to the rotation axis.
Sections~\ref{sec:case}, \ref{sec:3d} and \ref{sec:leveque} verify that
the base phase avoids the centre or axis in each reported case.

Section~\ref{sec:step2} obtains the laboratory velocity by differentiating
\eqref{eq:idea}. For rotation about $\bc$ this yields
\begin{equation}
  \bv(t,\bx) = \dot\theta\,(\bx - \bc)^{\perp}
     + \bQ\bigl(\theta(t)\bigr)\,
       \bvh\Bigl(t,\; \bc + \bQ\bigl(\theta(t)\bigr)^{\!\top}(\bx - \bc)\Bigr),
  \label{eq:videa}
\end{equation}
with $(r_1,r_2)^{\perp} = (-r_2,r_1)$: the velocity of the rotation plus the
base velocity. The base velocity is
evaluated at the point $\by = \bc + \bQ^{\!\top}(\bx-\bc)$ of the base flow that
the rotation moves to $\bx$, and it is turned by the same rotation.

The rigid motion can also be read as a moving coordinate frame: $\by$ is the
position relative to a frame that rotates with $E_t$, and $\bx = E_t(\by)$ the
position in the laboratory. Section~\ref{sec:step2} uses this view and allows a
frame with $E_0 \neq \mathrm{Id}$, for which $E_T = E_0$ suffices.

The base return and the returning frame give the exact endpoint. The
pointwise separation \eqref{eq:mirroridea} tests whether mirrored-time
retracing is broken. Sections~\ref{sec:step1}--\ref{sec:div} establish these
properties and the preservation of incompressibility needed for the
benchmarks below.

For the analytic illustration in Figure~\ref{fig:points},
take a disc of radius $0.26$ centred at $(0.55,0)$ and use
\begin{equation}
  \by(t)=
    \begin{pmatrix}1&k(t)\\0&1\end{pmatrix}\by_0,
  \qquad k(t)=1.6\sin^2(\pi t/T),
  \qquad \bx(t)=\bQ(2\pi t/T)\by(t).
  \label{eq:illustration}
\end{equation}
The base shear has determinant one, reverses after $T/2$, and returns
each label at $T$. The full turn is about $\bc=\mathbf0$, visibly separated
from the initial disc. These configurations are drawn directly from
\eqref{eq:illustration}. They illustrate the construction without adding
a numerical benchmark. The phase stays off the rotation centre: throughout
the base motion its first coordinate is at least
$0.55-0.26\sqrt{1+1.6^2}>0.059$. The winding argument therefore applies.

\subsection{Exact return of the base flow}
\label{sec:step1}

Let $a=T/2$. Choose a field $\bu(t,\by)$ that admits absolutely continuous
first-half trajectories on $[0,a]$ for the initial positions under
consideration. It may deform the initial phase strongly. To return the
material, apply the first-half field in reverse temporal order with the
opposite sign,
\begin{equation}
  \bvh(t,\by) =
  \begin{cases}
    \bu(t,\by),        & 0 \leq t \leq a, \\[2pt]
    -\bu(T-t,\by),     & a < t \leq T,
  \end{cases}
  \label{eq:vdef}
\end{equation}
or, written with the time $\tau$ measured from $a$,
\begin{equation}
  \bvh(a+\tau, \by) = -\,\bvh(a-\tau, \by), \qquad 0 < \tau \leq a .
  \label{eq:vantisym}
\end{equation}
Its flow map is denoted by $\Phih_s^t$, and the trajectory that starts at an
arbitrary $\by_0$ by $\by(t) = \Phih_0^t(\by_0)$. The value of $\bvh$ at the
single instant $t = a$ does not affect the trajectories, so $\bvh$ may jump
there when $\bu(a,\cdot) \neq 0$.

To show $\Phih_0^T=\mathrm{Id}$ on admissible initial positions, compare
the two halves as in Appendix~\ref{app:proof}. For $0 \leq \tau \leq a$,
view the positions at $a-\tau$ and $a+\tau$ as functions of $\tau$.
The chain rule, \eqref{eq:vdef} and $T-a = a$ give
    \begin{equation*}
      \frac{\dd}{\dd\tau}\,\by(a-\tau) = -\bu\bigl(a-\tau, \by(a-\tau)\bigr),
      \qquad
      \frac{\dd}{\dd\tau}\,\by(a+\tau) = -\bu\bigl(a-\tau, \by(a+\tau)\bigr),
    \end{equation*}
    so both functions satisfy
    \begin{equation}
      \frac{\dd\mathbf{w}}{\dd\tau} = -\bu(a-\tau, \mathbf{w}) .
      \label{eq:sameode}
    \end{equation}
At $\tau = 0$ both functions equal $\by(a)$. For uniqueness, assume
$\bu$ satisfies a spatial Lipschitz estimate for almost every $t$,
    \begin{equation}
      \lVert \bu(t,\by_1) - \bu(t,\by_2) \rVert \leq L(t)\, \lVert \by_1-\by_2 \rVert,
      \qquad \int_0^a L(t)\,\dd t < \infty ,
      \label{eq:lip}
    \end{equation}
    the squared distance $\lVert\boldsymbol\delta\rVert^2$ between two
    solutions of \eqref{eq:sameode} satisfies
    $\frac{\dd}{\dd\tau}\lVert\boldsymbol\delta\rVert^2 \leq
    2L(a-\tau)\lVert\boldsymbol\delta\rVert^2$. Gr\"{o}nwall's inequality keeps
    it at zero if it is zero initially.
Thus the two trajectories coincide,
    \begin{equation}
      \by(a+\tau) = \by(a-\tau), \qquad 0 \leq \tau \leq a ,
      \label{eq:retrace}
    \end{equation}
At $\tau = a$ this gives $\by(T) = \by(0) = \by_0$. As $\by_0$ was any
admissible initial position, $\Phih_0^T = \mathrm{Id}$.

The argument assumes existence of the first-half trajectories and uniqueness
of the reflected initial-value problem. Reflection of the first-half solution
itself supplies existence on the second half. Appendix~\ref{app:proof} gives
sufficient regularity assumptions. If uniqueness fails, trajectories can
merge and the reflected field need not prescribe a unique return. In one dimension,
$u(y) = -\operatorname{sign}(y)\sqrt{\lvert y\rvert}$ brings every point with
$0 < y_0 \leq a^2/4$ to rest at $y=0$ by $t=a$. The reversed field
$\operatorname{sign}(y)\sqrt{\lvert y\rvert}$ admits both the solution that stays
at $0$ and solutions that leave $0$ at any time, so $y(T)$ is not determined by
$y_0$. For the benchmark, the ambient field must be defined along the
required trajectories and retain the needed regularity and incompressibility.
Section~\ref{sec:domain} gives an extension and a confinement check for the
vortex.

Equation~\eqref{eq:retrace} specifies the temporal reversal that the cyclic
frame can break while preserving the exact endpoint.

\subsection{Laboratory velocity and trajectory separation}
\label{sec:step2}

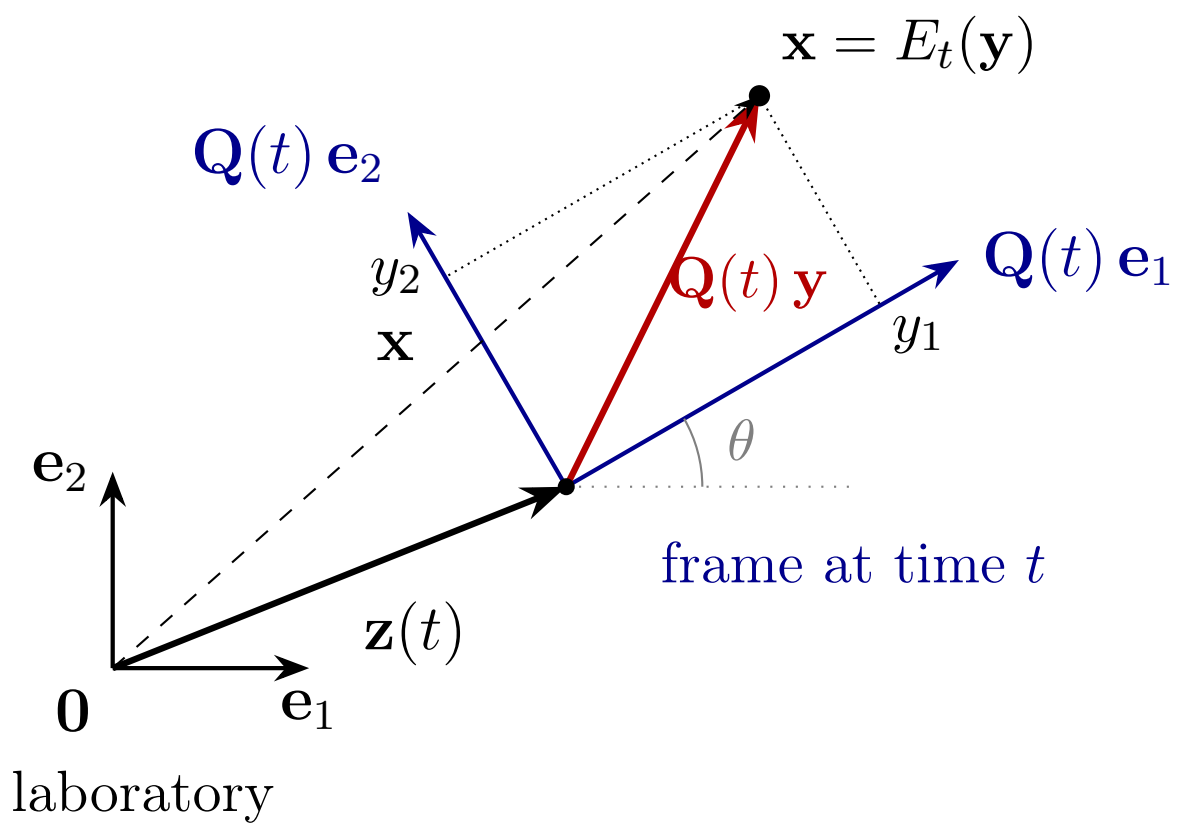
\begin{figure}[t]
\centering
\begin{tikzpicture}[>=Stealth, scale=1.0]
  \coordinate (O) at (0,0);
  \coordinate (z) at (3,1.2);
  \coordinate (x) at (4.278,3.786);
  \coordinate (y1) at (5.078,2.4);
  \coordinate (y2) at (2.2,2.586);
  \draw[->, thick] (O) -- (1.3,0) node[below] {$\be_1$};
  \draw[->, thick] (O) -- (0,1.3) node[left] {$\be_2$};
  \node[below left] at (O) {$\mathbf{0}$};
  \node[font=\small, align=left] at (0.2,-0.85) {laboratory};
  \draw[dotted, gray] (z) -- ++(1.9,0);
  \draw[->, thick, blue!55!black] (z) -- ++(30:3.0) node[right] {$\bQ(t)\,\be_1$};
  \draw[->, thick, blue!55!black] (z) -- ++(120:2.1) node[above left] {$\bQ(t)\,\be_2$};
  \draw[gray] (z) ++(0:0.9) arc (0:30:0.9);
  \node[gray, font=\small] at ($(z)+(15:1.2)$) {$\theta$};
  \node[font=\small, blue!55!black, align=center] at (4.9,0.7) {frame at time $t$};
  \draw[->, very thick] (O) -- (z) node[midway, below right] {$\bz(t)$};
  \draw[->, very thick, red!70!black] (z) -- (x);
  \node[red!70!black, font=\small] at (4.2,2.55) {$\bQ(t)\,\by$};
  \draw[->, dashed] (O) -- (x) node[midway, above left] {$\bx$};
  \draw[densely dotted] (x) -- (y1) node[pos=1, below right=-2pt, font=\small] {$y_1$};
  \draw[densely dotted] (x) -- (y2) node[pos=1, left, font=\small] {$y_2$};
  \fill (x) circle (2pt);
  \node[above right, font=\small] at (x) {$\bx = E_t(\by)$};
  \fill (z) circle (1.6pt);
\end{tikzpicture}
\caption{The rigid frame at time $t$: its origin is at $\bz(t)$ and its axes are
  rotated by $\bQ(t)$, here by the angle $\theta(t)$. A point with coordinates
  $\by = (y_1, y_2)$ relative to the frame has the laboratory position
  $\bx = E_t(\by) = \bz(t) + \bQ(t)\,\by$. The frame returns at $T$ if
  $\bz(T) = \bz(0)$ and $\bQ(T) = \bQ(0)$.}
\label{fig:Esketch}
\end{figure}

A moving rigid frame is described by the position $\bz(t)$ of its origin and its
orientation, a rotation matrix $\bQ(t)$ with $\bQ(t)\bQ(t)^{\!\top}=\bI$
and $\det\bQ(t)=1$. Assume $\bz$ and $\bQ$ are continuously differentiable.
A point with coordinates $\by$ relative to the frame has the laboratory
position (Fig.~\ref{fig:Esketch})
\begin{equation}
  \bx = E_t(\by) = \bz(t) + \bQ(t)\,\by .
  \label{eq:E}
\end{equation}

Let the material move with the base field relative to the frame,
$\by(t) = \Phih_0^t(\by_0)$, and follow it in the laboratory,
$\bx(t) = E_t\bigl(\by(t)\bigr)$. Differentiating with the product rule and
inserting $\by = \bQ^{\!\top}(\bx - \bz)$ gives the velocity of the point in the
laboratory,
\begin{equation}
  \dot \bx = \dot \bz + \dot \bQ\,\by + \bQ\,\dot \by
         = \dot \bz + \dot \bQ \bQ^{\!\top}(\bx - \bz)
           + \bQ\,\bvh\bigl(t, \bQ^{\!\top}(\bx - \bz)\bigr).
\end{equation}
The right-hand side defines the laboratory field
\begin{equation}
  \bv(t,\bx) = \dot \bz(t) + \dot \bQ(t)\bQ(t)^{\!\top}\bigl(\bx - \bz(t)\bigr)
         + \bQ(t)\, \bvh\bigl(t,\; \bQ(t)^{\!\top}(\bx-\bz(t))\bigr),
  \label{eq:w}
\end{equation}
the transformation law of a velocity field under a superposed rigid motion. Its
three terms are the translation of the frame, its rotation, and the base
velocity, evaluated at the position relative to the frame and rotated into the
laboratory. The laboratory flow is defined wherever the corresponding base
trajectory exists. Uniqueness follows by applying the inverse frame to any
laboratory solution. With $\by_0=E_0^{-1}(\bx_0)$, the laboratory trajectory is
\begin{equation}
  \Phi_0^t(\bx_0) = E_t\Bigl(\Phih_0^t\bigl(E_0^{-1}(\bx_0)\bigr)\Bigr).
  \label{eq:conj}
\end{equation}
At $T$ the base flow map is the identity, and \eqref{eq:conj} reduces to
$\Phi_0^T(\bx_0) = E_T\bigl(E_0^{-1}(\bx_0)\bigr)$. Thus each admissible
point returns if the frame closes,
\begin{equation}
  \bz(0) = \bz(T), \qquad \bQ(0) = \bQ(T) .
  \label{eq:frameperiodic}
\end{equation}

Only the endpoint poses enter this one-cycle return criterion. When $E_0$
differs from the identity, the base initial phase is $E_0^{-1}(\Op(0))$.

By \eqref{eq:retrace}, $\by(a+\tau)=\by(a-\tau)$. The laboratory separation
at the mirrored times is
\begin{equation}
  \bx(a+\tau) - \bx(a-\tau)
  = \bigl(\bz(a+\tau)-\bz(a-\tau)\bigr)
  + \bigl(\bQ(a+\tau)-\bQ(a-\tau)\bigr)\, \by(a-\tau) .
  \label{eq:noretrace}
\end{equation}
With $t=a-\tau$, the separation vanishes precisely when the relative
motion $E_t^{-1}E_{T-t}$ fixes $\by(t)$. Identical poses give coincidence
for all labels, while distinct poses can still coincide on their relative
fixed set. For a full turn about a fixed axis in three dimensions,
\eqref{eq:gapidea} holds with distance to the axis in place of distance to
the planar centre. The translation \eqref{eq:z1d} is the case $d=1$ and
$\bQ=1$.

For the interface $\Gamma(t)=\partial\Op(t)$, the relation
$\Gamma(T-t)=E_{T-t}E_t^{-1}\Gamma(t)$ shows that mirrored shapes are
congruent. Shape symmetry can make the contour sets coincide even when
individual labels occupy different positions.

\subsection{Incompressibility}
\label{sec:div}

The results use incompressible advection, with discrete carrier-volume
balance addressed separately in Section~\ref{sec:constraint}. For a
spatially Lipschitz base field with $\nabla_{\by}\cdot\bvh=0$ almost
everywhere, each term of \eqref{eq:w} is divergence-free. The spatial
derivatives below hold almost everywhere and also give zero distributional
divergence. The translation $\dot \bz(t)$ does
not depend on $\bx$. The rotation term has the gradient $\dot \bQ\bQ^{\!\top}$,
which is skew-symmetric because $\bQ\bQ^{\!\top} = \bI$, so its trace vanishes.
The rotated base velocity has the divergence
\begin{equation}
  \nabla_{\bx}\cdot \bQ\,\bvh\bigl(t,\bQ^{\!\top}(\bx-\bz)\bigr)
  = \operatorname{tr}\bigl(\bQ(\nabla_{\by} \bvh)\bQ^{\!\top}\bigr)
  = \operatorname{tr}(\nabla_{\by} \bvh) = 0 .
\end{equation}
The same gradient calculation shows that the symmetric part of
$\nabla_{\bx}\bv$ is the rotated symmetric part of
$\nabla_{\by}\bvh$. The added rotation gradient is skew-symmetric. This is
the local rate-of-deformation counterpart of the rigid correspondence between
the exact phases. The endpoint argument itself uses only invertibility and
$E_T=E_0$.

\subsection{Conditions for the benchmark construction}
\label{sec:requirements}

Table~\ref{tab:requirements} gives the conditions for exact return,
mirrored-time separation and incompressibility. Section~\ref{sec:agnostic}
gives a mesh algorithm.

\begin{table}[t]
\centering
\caption{Requirements of the construction.}
\label{tab:requirements}
\small
\begin{tabular}{@{}>{\raggedright\arraybackslash}p{0.25\textwidth}
                >{\raggedright\arraybackslash}p{0.36\textwidth}
                >{\raggedright\arraybackslash}p{0.31\textwidth}@{}}
\toprule
Property & Why it is needed & How it is obtained\\
\midrule
Base flow returns each admissible point
  & $\Op(T) = \Op(0)$ is the exact solution
  & reversed second half \eqref{eq:vdef}, with existence and uniqueness
    as in Appendix~\ref{app:proof}\\[2pt]
Frame returns at $T$
  & initial and final base positions have the same laboratory placement
  & $\bz(0)=\bz(T)$, $\bQ(0)=\bQ(T)$ \eqref{eq:frameperiodic}\\[2pt]
Mirrored-time separation
  & breaks the exact retracing symmetry when the same material point
    has different positions at a mirrored pair
  & choose and check the relative frame motion using \eqref{eq:noretrace}\\[2pt]
Divergence-free laboratory field
  & preserves incompressibility at the continuum level
  & $\nabla\cdot \bu = 0$ and a rigid frame, Section~\ref{sec:div}\\
\bottomrule
\end{tabular}
\end{table}

\subsection{Explicit rotating-frame formulas}
\label{sec:explicit}

For a planar rotation with angle $\theta(t)$,
\begin{equation}
  \bQ(t) = \begin{pmatrix} \cos\theta(t) & -\sin\theta(t)\\
                         \sin\theta(t) & \phantom{-}\cos\theta(t)\end{pmatrix},
  \qquad
  \dot \bQ(t)\bQ(t)^{\!\top} = \dot\theta(t)\begin{pmatrix}0&-1\\1&0\end{pmatrix},
\end{equation}
so that in two dimensions
\begin{equation}
  \bv(t,\bx) = \dot \bz(t) + \dot\theta(t)\bigl(\bx-\bz(t)\bigr)^{\!\perp}
         + \bQ(t)\,\bvh\bigl(t,\bQ(t)^{\!\top}(\bx-\bz(t))\bigr),
  \label{eq:w2d}
\end{equation}
with $(r_1,r_2)^{\perp} = (-r_2,r_1)$.
In three dimensions, a rotation about the first coordinate axis is
\begin{equation}
  \bQ(t) = \begin{pmatrix}
    1 & 0 & 0\\
    0 & \cos\theta(t) & -\sin\theta(t)\\
    0 & \sin\theta(t) & \phantom{-}\cos\theta(t)
  \end{pmatrix},
  \qquad
  \dot \bQ(t)\bQ(t)^{\!\top} = \dot\theta(t)\begin{pmatrix}
    0&0&0\\ 0&0&-1\\ 0&1&0 \end{pmatrix}.
\end{equation}
The special case $\bz(t)\equiv\mathbf{0}$, $\theta(t)=t$ gives a pure rotating frame; it is
closed over $[0,T]$ when $T=2\pi$ in this example. In general the planar
rotation closes when $\theta(T)-\theta(0)$ is an integer multiple of $2\pi$.

\subsection{Base vortex on the unit square}
\label{sec:basevortex}

One standard divergence-free deformation pattern on $[0,1]^2$ is generated by the
stream function
\begin{equation}
  \psi_0(x_1,x_2) = \frac{1}{\pi}\sin^2(\pi x_1)\sin^2(\pi x_2),
  \label{eq:psi0}
\end{equation}
which gives
\begin{equation}
  \bu_0(x_1,x_2) = \begin{pmatrix} -\sin^2(\pi x_1)\sin(2\pi x_2)\\[2pt]
                                  \phantom{-}\sin(2\pi x_1)\sin^2(\pi x_2)
                 \end{pmatrix}.
  \label{eq:u0}
\end{equation}
Multiplying \eqref{eq:u0} by an arbitrary integrable amplitude $\gamma(t)$ on
$[0,a]$ gives a time-dependent first-half field $\bu(t,\by)=\gamma(t)\bu_0(\by)$.
Combining the reflected base field with the cyclic frame gives a
divergence-free laboratory field with exact return. In the base coordinates the solution follows the autonomous
flow of $\bu_0$ at signed flow time $s(t)=\int_0^t\gamma(\tau)\,\dd\tau$
on the first half, with the reflected amplitude on the second half.

\subsection{Volume fraction transport and error measurement}
\label{sec:vofeq}

The phase is described by its indicator $\chi(t,\bx)$, equal to $1$ in
$\Op(t)$ and $0$ elsewhere. A geometric VOF method decomposes $\Omega$ into
cells $\Omega_k$, $k \in K$, and advects their volume fractions
$\alpha_k(t) = |\Omega_k|^{-1}\int_{\Omega_k}\chi(t,\bx)\,\dd V$
\cite{maric2020}. For a solenoidal velocity, the integral form of the transport
equation for $\chi$, integrated over a time step $[t^n, t^{n+1}]$, gives
\begin{equation}
  \alpha_k(t^{n+1}) = \alpha_k(t^n) - \frac{1}{|\Omega_k|}
    \int_{t^n}^{t^{n+1}}\!\!\int_{\partial\Omega_k} \chi\,\bv\cdot\bn \,\dd S\,\dd t ,
  \label{eq:vof}
\end{equation}
an exact equation, which the geometric VOF method approximates
\cite{maric2020}. Wherever the backward ambient flow is defined,
$\chi(t,\bx)=\chi(0,\Phi_t^0(\bx))$. Exact return of the phase gives
$\chi(T,\cdot) = \chi(0,\cdot)$, so the exact volume fractions at $T$ are the
initial ones, $\alpha_k(T) = \alpha_k(0)$. This common reference permits comparison of the paired runs by volume
error, symmetric difference, interface distance or shape moments.

\section{Solenoidal face fluxes for geometric VOF}
\label{sec:vof}

\subsection{Cell balance from edge circulation}
\label{sec:constraint}

Geometric VOF reconstructs an interface within each interfacial cell and
approximates the phase volume transported through each face over the time
step \cite{maric2020}. The volumetric carrier fluxes $F_f$ should satisfy
cell-wise balance,

\begin{equation}
  \sum_{f\in F_k} F_f = 0 \qquad\text{for every cell } \Omega_k ,
  \label{eq:discretediv}
\end{equation}
where $F_k$ is the set of faces of $\Omega_k$, and the fluxes are taken with the
outward orientation.

A face-centre evaluation or interpolation of a divergence-free field can
leave a cell flux imbalance. Cell-wise carrier balance is the discrete
counterpart of incompressibility. It does not by itself ensure accurate or bounded phase
transport. In the global phase balance, shared conservative phase
fluxes cancel between cells, leaving boundary transport and any sources
or numerical corrections.

For consistently oriented closed cell boundaries, \eqref{eq:discretediv}
holds in exact arithmetic if the fluxes are
constructed as a discrete circulation of a vector potential. In two dimensions,
embedded in the ordered $x$--$z$ plane, use $\bA = \psi\,\be_y$, so that
$\nabla\times\bA=(-\partial_z\psi,0,\partial_x\psi)$. Then
\begin{equation}
  \widehat F_f = \oint_{\partial S_f}\psi\,\dd y, \qquad
  F_f =
  \sum_{e=(p_1,p_2)\in\partial S_f}
  \tfrac12\bigl(\psi_{p_1}+\psi_{p_2}\bigr)\bigl(y_{p_2}-y_{p_1}\bigr),
  \label{eq:circ}
\end{equation}
where $\widehat F_f$ is the exact instantaneous flux and $F_f$ is its
edge-trapezoid approximation. The sum uses vertex samples of $\psi$ and
oriented edges of the face polygon. When the faces of one cell are traversed with their
outward orientation, every edge of the cell is traversed exactly twice in
opposite senses, so the contributions cancel identically. This is the discrete
counterpart of $\nabla\cdot(\nabla\times \bA)=0$. Adding a smooth gradient
to the potential leaves the exact circulation invariant. Its edge
quadrature can change, while cell-wise cancellation is preserved.

\begin{remark}[Design rule]
\label{rem:rule}
For this flux assembly, express a modification of the test field as a
modification of the potential at mesh vertices and reuse the oriented
circulation.
\end{remark}

\subsection{Stream function of the laboratory-frame field}
\label{sec:streamfun}

To apply Remark~\ref{rem:rule} to Section~\ref{sec:method}, we need the
laboratory-frame potential. In two dimensions write $\bJ$ for rotation by
$+\pi/2$ and use the convention in \eqref{eq:psi0}--\eqref{eq:u0},
\begin{equation}
  \bJ = \begin{pmatrix}0&-1\\1&0\end{pmatrix},
  \qquad
  \bu = \bJ\,\nabla\psi .
  \label{eq:convention}
\end{equation}
For the explicit benchmark formulas, specialise the base field to
$\bvh(t,\bxi)=\gamma(t)\bu_0(\bxi)$, where $\gamma$ now denotes the
antisymmetric amplitude on the full interval $[0,T]$. Let the frame rotate
by $\theta(t)$ about a fixed centre $\bc$ and translate by
$\boeta(t)$, i.e. $E_t(\bxi) = \boeta(t) + \bc + \bQ(t)(\bxi-\bc)$, and set
$\br := \bx-\boeta(t)-\bc$ and $\bxi := \bc + \bQ(t)^{\!\top}\br$. The
laboratory-frame field \eqref{eq:w2d} is then
\begin{equation}
  \bv(t,\bx) = \dot{\boeta}(t) + \dot\theta(t)\,\br^{\perp}
         + \bQ(t)\,\bvh(t,\bxi), \qquad \bvh(t,\bxi) = \gamma(t)\bu_0(\bxi).
  \label{eq:wframe}
\end{equation}

\begin{proposition}[Stream function of the laboratory-frame field]
\label{prop:psi}
The field \eqref{eq:wframe} satisfies $\bv = \bJ\nabla_{\bx}\Psi$ with
\begin{equation}
  \Psi(t,\bx) =
  \underbrace{\dot{\eta}_2(t)\,x_1 - \dot{\eta}_1(t)\,x_2}_{\text{translation}}
  \;+\;
  \underbrace{\tfrac12\,\dot\theta(t)\,\lVert \br\rVert^{2}}_{\text{rigid rotation}}
  \;+\;
  \underbrace{\gamma(t)\,\psi_0(\bxi)}_{\text{base vortex}} .
  \label{eq:Psi}
\end{equation}
\end{proposition}

\begin{proof}
The three terms are treated separately. For the translation,
the gradient of the first term is $(\dot{\eta}_2,-\dot{\eta}_1)^{\top}$,
which gives $\dot{\boeta}$ after multiplication by $\bJ$. For the rotation,
$\nabla_{\bx}\bigl(\tfrac12\dot\theta\lVert \br\rVert^2\bigr) = \dot\theta\,\br$,
hence $\bJ\dot\theta \br = \dot\theta \br^{\perp}$. For the base term, note that
$\partial\xi_i/\partial x_j = (\bQ^{\!\top})_{ij}$, so the chain rule gives
$\nabla_{\bx}(\psi_0\circ\bxi) = \bQ\nabla_{\bxi}\psi_0$ and therefore
\begin{equation}
  \bJ\nabla_{\bx}(\psi_0\circ\bxi) = \bJ\bQ\nabla_{\bxi}\psi_0
  = \bQ\bJ\nabla_{\bxi}\psi_0 = \bQ\,\bu_0(\bxi),
\end{equation}
where $\bJ\bQ=\bQ\bJ$ because $\bJ$ and $\bQ$ both lie in the abelian group
$\SOgroup(2)$.
\end{proof}

Because rotations commute in $\SOgroup(2)$, the base contribution uses
$\psi_0$ at the pulled-back point $\bxi$ with the same time factor $\gamma$.
Substituting \eqref{eq:Psi} for $\psi$ in \eqref{eq:circ} gives the full
laboratory flux. For any admissible frame the edge cancellation enforces
\eqref{eq:discretediv} in exact arithmetic.

In three dimensions the same route applies with a vector potential $\bA$ and
$\widehat F_f = \oint_{\partial S_f} \bA\cdot\dd \mathbf{l}$, using
\begin{equation}
  \bA_{\mathrm{t}} = \tfrac12\,\dot \bz\times \bx, \qquad
  \bA_{\mathrm{r}} = -\tfrac12\lVert \br_{\perp}\rVert^{2}\,\bomega, \qquad
  \bA_{\mathrm{base}}(\bx) = \gamma(t)\,\bQ\,\bA_0\bigl(\bQ^{\!\top}(\bx-\bz)\bigr),
\end{equation}
where $\br=\bx-\bz(t)$, $\bomega$ is the frame angular velocity and
$\br_\perp$ is the component of $\br$ orthogonal to $\bomega$. The rotation
term is zero when $\bomega=\mathbf{0}$. The base term follows from equivariance
of the curl under proper rotations for the base potential $\gamma(t)\bA_0$.
In the ordered $x$--$z$ plane the embedding $\bA=\Psi\be_y$ gives
$(-\partial_z\Psi,0,\partial_x\Psi)$. The horizontal part of the shear flow
in Section~\ref{sec:3d} uses $\psi_0\be_z$, which gives
$(\partial_y\psi_0,-\partial_x\psi_0,0)$.

\subsection{Confinement of the material support and domain sizing}
\label{sec:domain}

The fixed mesh must contain the swept phase over the entire cycle.
For the vortex, a streamline invariant bounds the support and gives
sufficient domain dimensions for a chosen rotation centre.

\begin{proposition}[Invariance of the streamline band]
\label{prop:band}
Let $\bvh(t,\bxi)=\gamma(t)\bu_0(\bxi)$ with $\bu_0 = \bJ\nabla\psi_0$. Then
$\psi_0$ is constant along base-flow trajectories.
\end{proposition}

\begin{proof}
$\tfrac{\dd}{\dd t}\psi_0(\bxi(t)) = \gamma(t)\,\nabla\psi_0\cdot \bJ\nabla\psi_0
= 0$, since $\bJ$ is skew-symmetric.
\end{proof}

For the disc in Section~\ref{sec:case}, $\psi_0$ lies between
$\sin^2(\pi/10)/\pi\simeq 0.030396$ and
$\sin^2(3\pi/5)/\pi\simeq 0.287914$. To see the lower bound, minimize at fixed
height by taking the largest allowed $|x-0.5|$. Write $y=0.75+b$ on the disc
with $b\in[-0.15,0.15]$ and set $q=\sqrt{0.15^2-b^2}$. The derivative of
$\log(\cos(\pi q)\sin(\pi(0.75+b)))$, divided by $\pi$, is
$b\tan(\pi q)/q+\cot(\pi(0.75+b))$, with the continuous endpoint
interpretation. For $b\le0$ both terms are nonpositive, and for $b\ge0$
the first is at most $\tan(0.15\pi)<1$ while the second is at most $-1$.
The minimum is attained at $(0.5,0.9)$. The upper bound is attained at
$(0.5,0.6)$ since each sine factor is maximal there among points in the disc.

Let $\bc_*=(0.5,0.5)$ and put $r=\|\bxi-\bc_*\|$ in the base square.
The function $g(s)=\log\cos(\pi\sqrt{s})$ is concave for $0\le s<1/4$,
since $\tan q/q$ increases for $0<q<\pi/2$. For $a_i=|\xi_i-0.5|$,
applying concavity to $a_1^2,a_2^2$ gives
\begin{equation}
  \pi\psi_0(\bxi) \leq \cos^4(\pi r/\sqrt{2}) .
  \label{eq:horizontalbound}
\end{equation}
The inequality is attained along the diagonals. Applying it to the lower
level of the streamline band gives the sharp centred radius
\[
  R_*=\frac{\sqrt{2}}{\pi}\arccos\sqrt{\sin(\pi/10)}
      =0.441766\ldots <0.442 .
\]
For a rotation about $\bc$ and an offset $\boeta(t)$ the laboratory phase
lies in the ball about $\bc$ of radius
$R_*+\|\bc-\bc_*\|+\sup_{[0,T]}\|\boeta\|$. Table~\ref{tab:domain} gives
conservative square side lengths for three centres and zero offset. The
centred rotation has a physical clearance of at least $0.058$, about $1.9$
cell widths at $N=32$ and $119$ at $N=2048$.

\begin{table}[htbp]
\centering
\small
\caption{Sufficient square side lengths for the vortex disc, using the centred
streamline-band radius and the triangle inequality. Each square is centred at
the rotation centre $\bc$ and there is no frame translation.}
\label{tab:domain}
\begin{tabular}{lcc}
\toprule
Rotation centre $\bc$ & Upper bound on support radius & Sufficient square side\\
\midrule
$(0.5,\,0.5)$  & $0.442$ & $0.884$ \\
$(0.5,\,0.75)$ & $0.692$ & $1.384$ \\
$(0.2,\,0.2)$  & $0.867$ & $1.734$ \\
\bottomrule
\end{tabular}
\end{table}

Where $\bxi$ leaves the unit square, $\psi_0$ is extended by zero. Since
$\partial_1\psi_0 = \sin(2\pi x_1)\sin^2(\pi x_2)$ and
$\partial_2\psi_0 = \sin^2(\pi x_1)\sin(2\pi x_2)$ both vanish on
$\partial[0,1]^2$, together with $\psi_0$ itself, this extension is $C^1$. Its piecewise bounded second derivatives and
continuous gradient give a Lipschitz base velocity as in \eqref{eq:lip}.
Outside the base square the velocity is that of the frame alone.

\section{Results}
\label{sec:results}

\subsection{Reversed vortex and numerical protocol}
\label{sec:case}

We use the two-dimensional reversed-vortex case and solver distributed with
\texttt{TwoPhaseFlow} \cite{scheufler2023}. The mesh covers the unit square
in the $x$--$z$ plane with $N\times N$ cells and one cell in the transverse
direction. We initialise the phase geometrically as a disc of radius $r=0.15$
centred at $(0.5,0.75)$.

The base field is \eqref{eq:u0} with stream function \eqref{eq:psi0}.
A smooth amplitude implements the forward--backward construction,
\begin{equation}
  \gamma(t) = \cos\!\left(\frac{2\pi t}{T_{\mathrm{per}}}\right),
  \qquad T_{\mathrm{per}} = 16, \qquad T = 8 ,
  \label{eq:alpha}
\end{equation}
The run covers half a cosine period. With $a=T/2$,
$\gamma(a+\tau) = -\sin(\pi\tau/8) = -\gamma(a-\tau)$, so the field satisfies
the antisymmetry \eqref{eq:vantisym} and the base trajectories retrace as
shown in Section~\ref{sec:step1}.

The superposed frame, denoted Frame~A below, is one full revolution about the
centre of the domain,
\begin{equation}
  \bc = (0.5,\,0.5), \qquad \theta(t) = \frac{2\pi t}{T}, \qquad
  \boeta \equiv \mathbf{0} ,
\end{equation}

so that $\bQ(0)=\bQ(T)=\bI$, as required by \eqref{eq:frameperiodic}.
The support bound in Table~\ref{tab:domain} places the swept phase strictly
inside the unit square. The stream-function value at $\bc$ exceeds those
in the phase's streamline band. Equation~\eqref{eq:gapidea} therefore gives
distinct positions for each phase label at interior mirrored times. The
extruded square has unit depth. For Frame~A, we assemble carrier fluxes on
boundary faces by circulation; the unframed vortex retains the reference
implementation's boundary-flux assembly. All three benchmarks prescribe
zero phase fraction on the physical outer boundary. The transverse faces
of the two-dimensional case use \texttt{empty} boundary conditions.

\label{sec:setup}
The advection scheme is isoAdvector \cite{roenby2016}, tested with the plicRDF
\cite{scheufler2019} and isoAlpha reconstructions. It uses face fluxes and
cell-centred velocities evaluated at time-step midpoints. The framed vortex fluxes
use \eqref{eq:Psi} and \eqref{eq:circ}. Within each pair of runs, the framed
and unframed cases use a common time grid. Let $h=1/N$ and let
$U_{\mathrm{est}}$ be the maximum of the framed $\|\bv\|_1$ over space and time
samples. This estimate is greater than the corresponding unframed estimate
in each benchmark. We round the number of steps up to a multiple of four,
\begin{equation}
  \Delta t_* = 0.2h/U_{\mathrm{est}}, \qquad
  m=4\left\lceil T/(4\Delta t_*)\right\rceil,\qquad
  \Delta t=T/m .
  \label{eq:dt}
\end{equation}
We use the $1$-norm because the cell Courant number
$\tfrac12\sum_{f\in F_k}|F_f|\,\Delta t/|\Omega_k|$ equals
$\|\bv\|_1\Delta t/h$ for a uniform velocity on a cubic cell.
The maximum interface Courant indicators reported by the solver round to
$0.200$ for the vortex and shear and $0.154$ for the fully three-dimensional
deformation \cite{bothe2026figshare}. The solver evaluates this indicator using
the current phase fraction and the face fluxes available before the next
midpoint update. The time-step formula uses the sampled speed, and the
logged values describe the solver's diagnostic on that grid.

Let $V_k=|\Omega_k|$ and let $\alpha_{k,h}^{\mathrm{ref}}$ be the reference
fraction obtained by cutting each cell using samples of the implicit initial
interface. We measure the terminal, cell-volume-weighted discrepancy from
this reference and the absolute change in transported phase volume,
\begin{align}
  E_{\text{shape}}(T) &=
    \sum_k V_k \big|\alpha_k^h(T)-\alpha_{k,h}^{\mathrm{ref}}\big|,\\
  E_{\text{mass}}(t) &=
    \left|\sum_k V_k\big(\alpha_k^h(t)-\alpha_k^h(0)\big)\right| .
\end{align}
The archived boundedness diagnostic $E_{\text{bound}}$ takes the maximum
over MPI ranks of $|\max(-\min_k\alpha_k^h,\max_k\alpha_k^h-1)|$, with cell extrema
computed separately on each rank. This diagnostic equals the largest
excursion outside $[0,1]$ if each rank with no violation contains a
pure-phase cell. The tables give terminal values. The implementation
checks also report maxima over stored outputs or solver iterations, as
specified in the run summaries.

The shape diagnostic compares cell averages with a discrete geometric
reference. It need not equal the volume of the symmetric difference
between reconstructed phases. We fit refinement exponents to
$E_{\text{shape}}(T)$.

\subsection{Implementation checks}
\label{sec:verification}

We check carrier-flux balance, the identity frame and the effect of rigid
rotation separately. Section~\ref{sec:leveque} reports the vector-potential
checks. The \texttt{TwoPhaseFlow} implementation and its workflows were
written with an AI coding assistant (see the declaration at the end of the
article); the checks in this section test the implementation regardless of
how it was written.

We monitor the discrete balance \eqref{eq:discretediv} at each step. The
largest value of $\max_k\lvert\sum_{f\in F_k} F_f\rvert/|\Omega_k|$ over the
framed vortex runs is $3.6\times10^{-15}$. The residual is zero in the
recorded frame-only runs. The largest stored phase-volume change in the
two-dimensional study is $1.3\times10^{-13}$, at $N=2048$ \cite{bothe2026figshare},
\path{reversed-vortex/results/courant_check.txt}.

For the identity control, we set $\theta\equiv0$ and $\boeta\equiv\mathbf{0}$
in the moving-frame implementation and compare with the original
retracing case. The two common outputs are at $t=4$ and $t=8$. At $N=128$,
the two plicRDF runs give $E_{\text{shape}}(T)=2.9631509711\times10^{-3}$
to the displayed precision, with relative difference $7.5\times10^{-14}$.
This also bounds the relative difference at the earlier output. The reported
discrepancy at $t=4$ measures agreement of the two outputs' diagnostics
against the initial disc \cite{bothe2026figshare},
\path{reversed-vortex/results/frame_identity_check.txt}.

Setting $\gamma\equiv0$ reduces the test to a solid-body revolution of the disc.
The plicRDF control gives $E_{\text{shape}}(T) = 7.680\times10^{-4}$,
$1.862\times10^{-4}$ and $4.593\times10^{-5}$ at $N=64$, $128$ and $256$.
The observed exponents between successive grids are $2.04$ and $2.02$
\cite{bothe2026figshare}, \path{reversed-vortex/results/frame_only_check.txt}.
These values characterise the discretisation of the isolated rigid motion.
The framed vortex combines rigid motion and deformation.

\subsection{Interface evolution}
\label{sec:interfaces}

\begin{figure}[htbp]
\centering
\includegraphics[width=\textwidth]{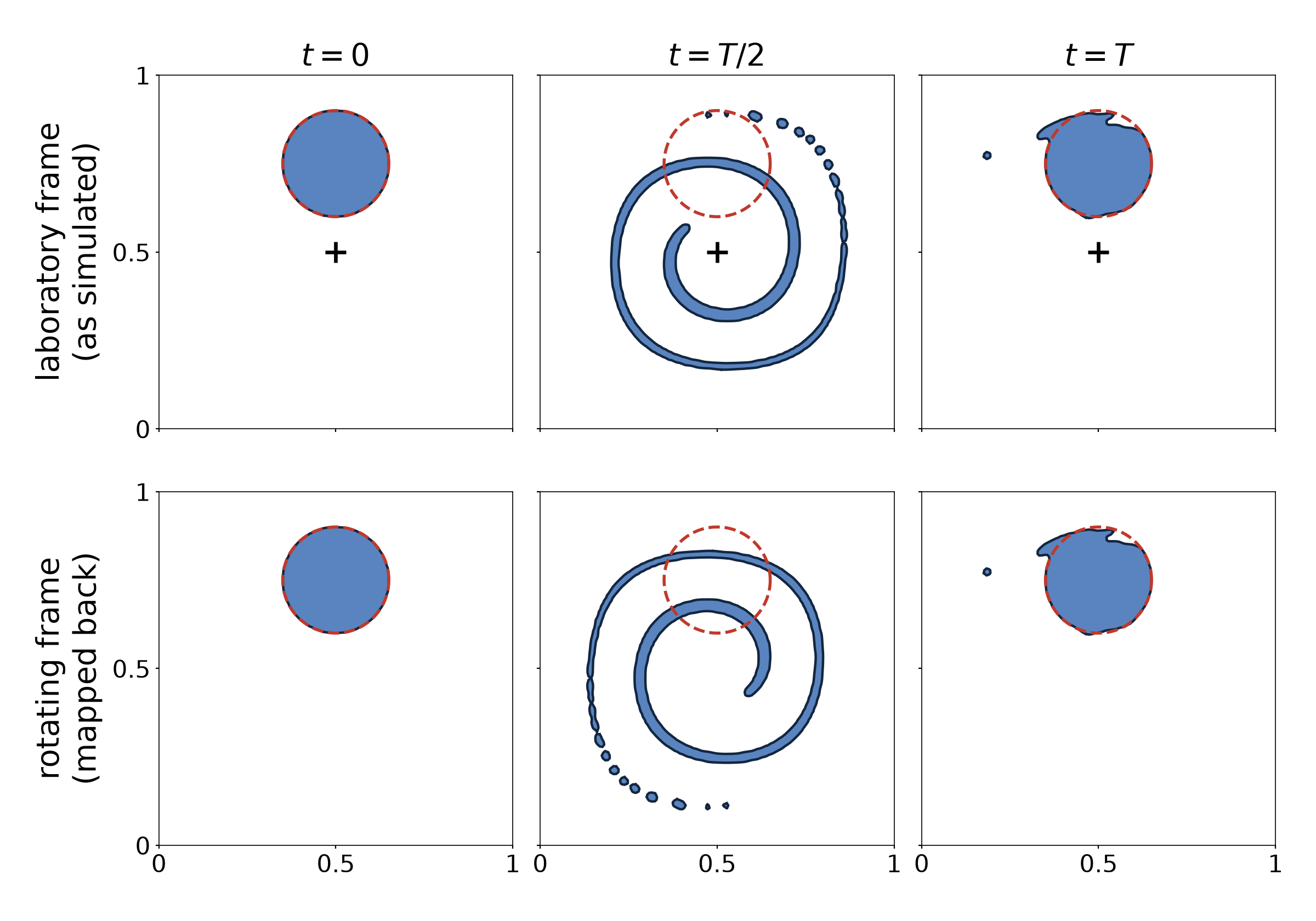}
\caption{Framed case (Frame~A), plicRDF, $N=128$, at $t=0$, $T/2$ and $T$. The top row shows the interface in laboratory coordinates. The bottom row
maps the same contours through $E_t^{-1}$ for comparison with
Figure~\ref{fig:static}. The rows coincide at $T$, where $E_T$ is the
identity. The dashed red circle gives the exact initial and final interface.
The $+$ symbol marks the rotation centre. Data in \cite{bothe2026figshare}, \protect\path{reversed-vortex/results/interface_moving_frame.png}.}
\label{fig:moving}
\end{figure}

\begin{figure}[htbp]
\centering
\includegraphics[width=\textwidth]{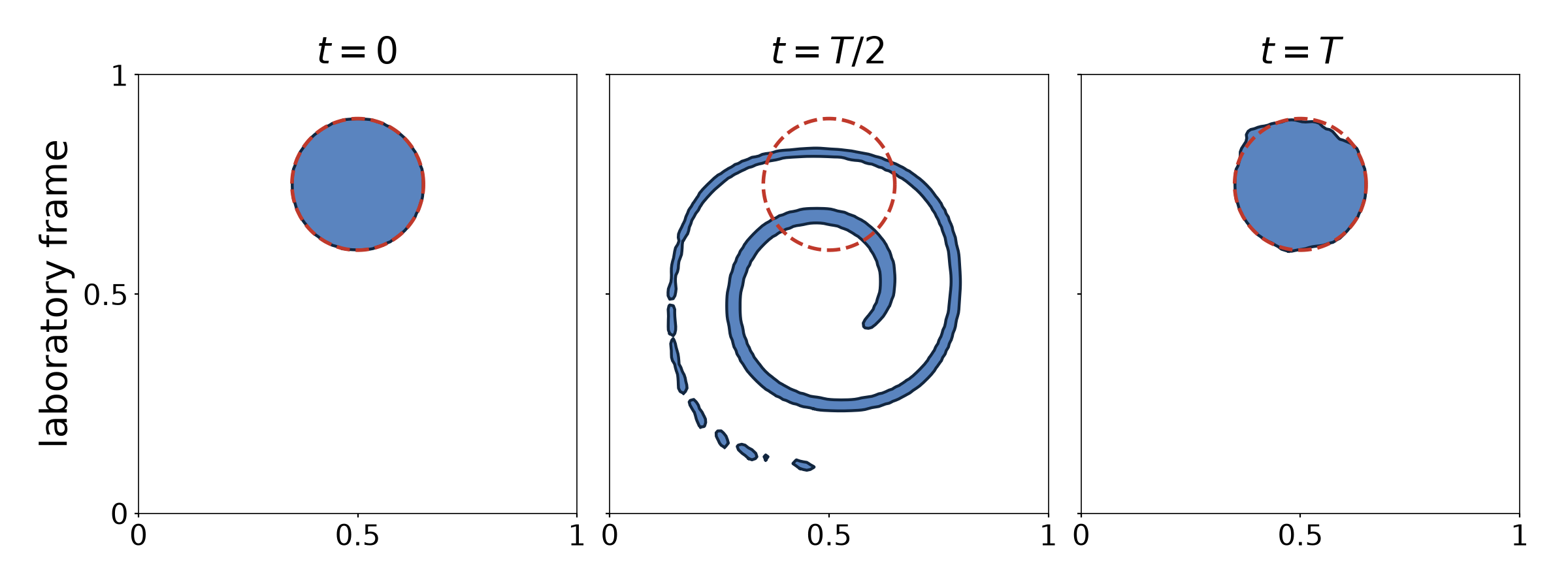}
\caption{Unframed case, plicRDF, $N=128$, at $t=0$, $T/2$ and $T$. Dashed red:
exact interface at $t=0$ and $t=T$. Data: \cite{bothe2026figshare},
\protect\path{reversed-vortex/results/interface_non_moving_frame.png}.}
\label{fig:static}
\end{figure}

Figures~\ref{fig:moving} and~\ref{fig:static} show the reconstructed
interfaces at $t=0$, $T/2$ and $T$ on the $N=128$ grid with plicRDF.
The snapshots come from separate visualisation runs with more frequent
outputs than the convergence runs.

At $T/2$ the numerical contours show fragments near the filament tip.
At $T$ fragments remain visible with the frame (Figure~\ref{fig:moving}).
The exact phase is the image of the initial disc under the flow map and
remains connected throughout the cycle.

\begin{figure}[htbp]
\centering
\includegraphics[width=\textwidth]{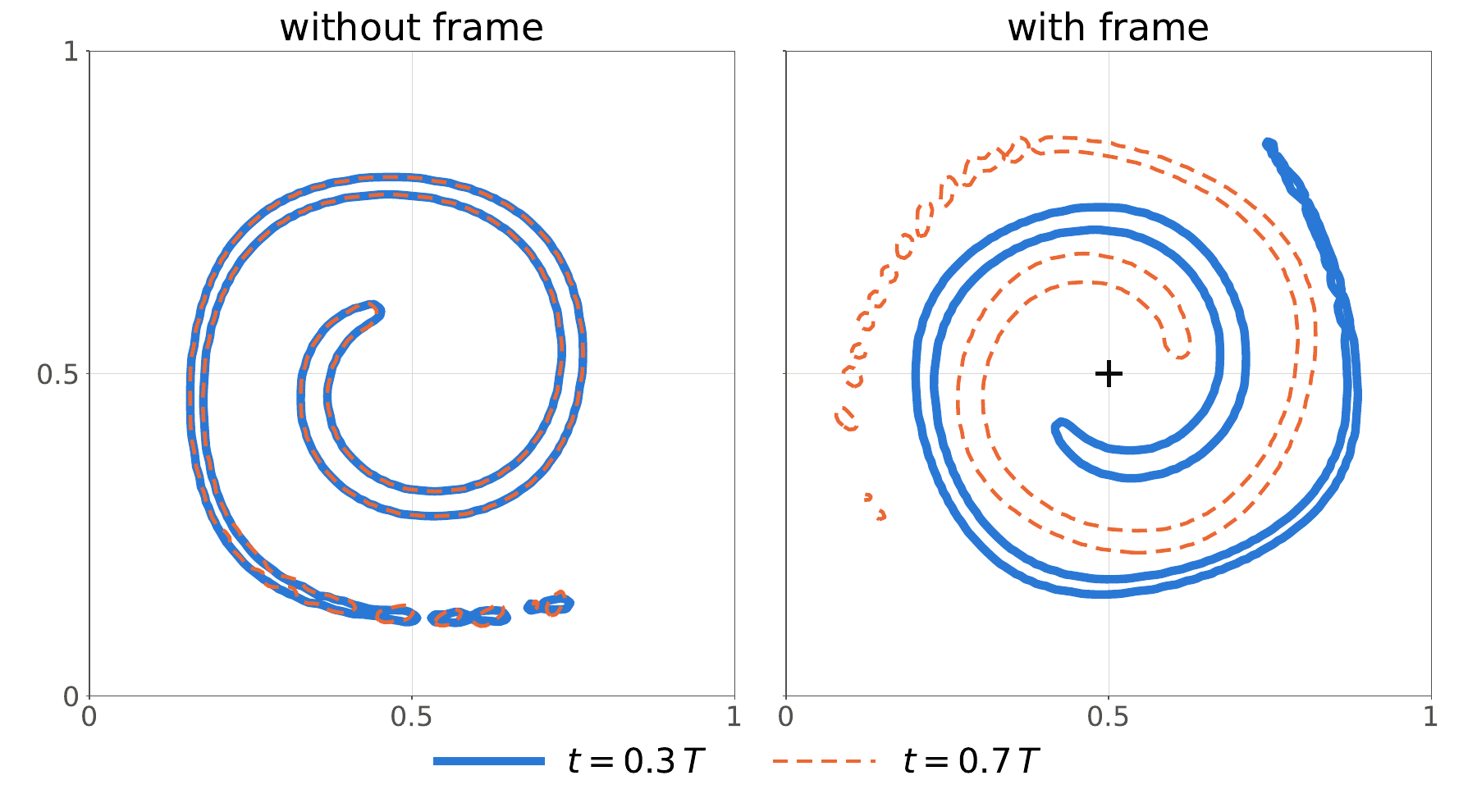}
\caption{Simulated interface ($\alpha=0.5$, plicRDF, $N=128$) at $t=0.3T$
(solid) and $0.7T$ (dashed). The unframed contours (left) nearly overlap at
plotting scale. The framed contours (right) occupy different locations.
Data in \cite{bothe2026figshare},
\protect\path{reversed-vortex/results/interface_evolution.pdf}; an animation of
both runs is \protect\path{reversed-vortex/results/interface_evolution.gif}.}
\label{fig:evolution}
\end{figure}

Figure~\ref{fig:evolution} compares the numerical $\alpha=0.5$ contours
at the mirrored times $0.3T$ and $0.7T$. Their relative placement follows
the pattern shown by the exact configurations: near overlap at plotting
scale without the frame and visible separation with Frame~A.

\subsection{Grid refinement}
\label{sec:convergence}

\begin{figure}[htbp]
\centering
\includegraphics[width=\textwidth]{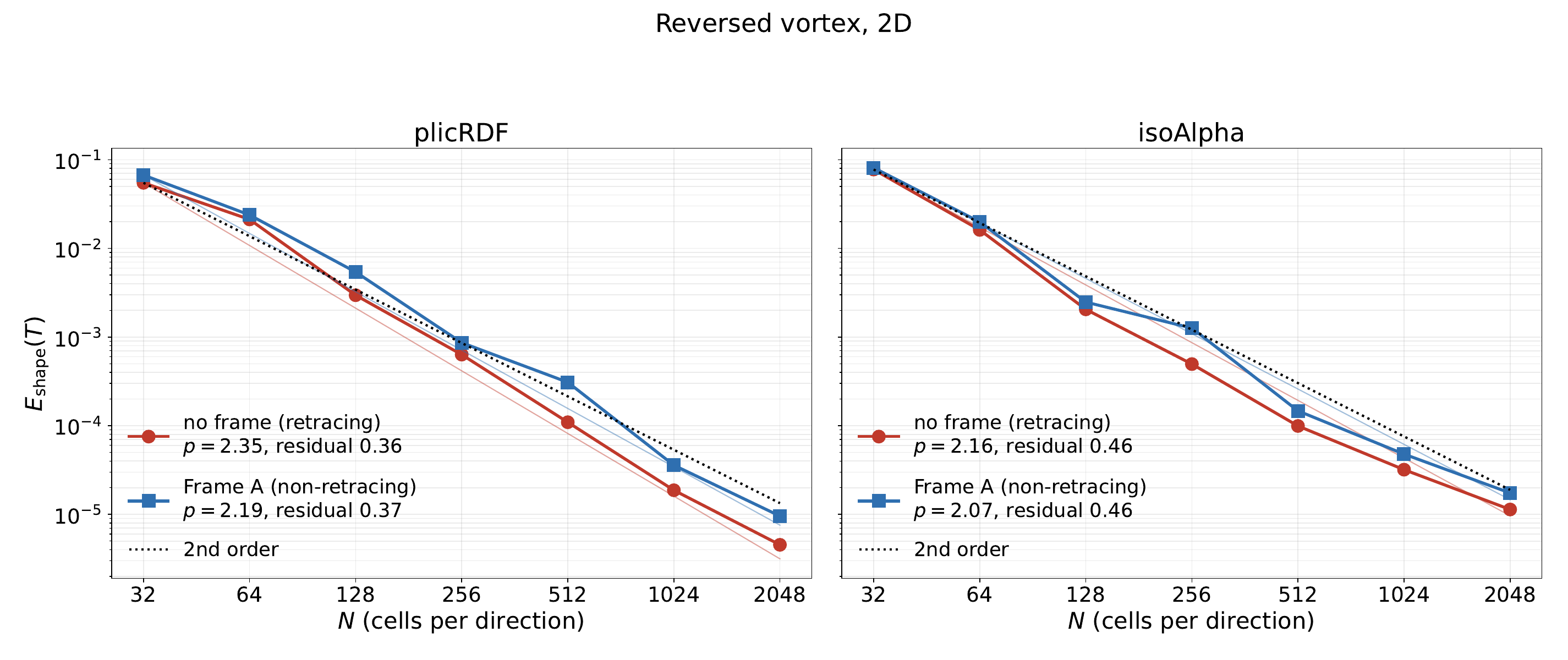}
\caption{Convergence of $E_{\text{shape}}(T)$ with and without the superposed
frame. The lines show least-squares fits on logarithmic axes. The legend gives
the fitted exponents and maximum absolute log residuals. Data: \cite{bothe2026figshare},
\protect\path{reversed-vortex/results/convergence_shape_error.png}.}
\label{fig:conv}
\end{figure}

Figure~\ref{fig:conv} and Table~\ref{tab:conv} show how the terminal
cell-weighted shape error changes with grid refinement.

\begin{table}[htbp]
\centering
\small
\setlength{\tabcolsep}{4pt}
\caption{Convergence on the reversed-vortex test, both reconstruction schemes. The order is the negative least-squares slope of $\ln E_{\text{shape}}$ against $\ln N$ over the whole sequence; the residual is the maximum absolute residual in $\ln E_{\text{shape}}$. Bold: the shape error with the frame exceeds the one without it at the same $N$. Data: \cite{bothe2026figshare}, \protect\path{reversed-vortex/results/convergence.csv}.}
\label{tab:conv}
\begin{tabular}{r cc cc}
\toprule
& \multicolumn{2}{c}{no frame (retracing)} & \multicolumn{2}{c}{Frame A (non-retracing)} \\
\cmidrule(lr){2-3}\cmidrule(lr){4-5}
$N$ & $E_{\text{shape}}$ & $E_{\text{bound}}$ & $E_{\text{shape}}$ & $E_{\text{bound}}$ \\
\midrule
\multicolumn{5}{l}{\emph{plicRDF}} \\
32 & $5.486\times10^{-2}$ & $0$ & {\boldmath$6.712\times10^{-2}$} & $8.470\times10^{-22}$ \\
64 & $2.120\times10^{-2}$ & $0$ & {\boldmath$2.381\times10^{-2}$} & $1.694\times10^{-21}$ \\
128 & $2.963\times10^{-3}$ & $0$ & {\boldmath$5.412\times10^{-3}$} & $1.131\times10^{-22}$ \\
256 & $6.338\times10^{-4}$ & $4.235\times10^{-22}$ & {\boldmath$8.643\times10^{-4}$} & $9.191\times10^{-8}$ \\
512 & $1.095\times10^{-4}$ & $5.672\times10^{-12}$ & {\boldmath$3.091\times10^{-4}$} & $1.804\times10^{-7}$ \\
1024 & $1.873\times10^{-5}$ & $6.557\times10^{-11}$ & {\boldmath$3.585\times10^{-5}$} & $1.053\times10^{-9}$ \\
2048 & $4.540\times10^{-6}$ & $4.482\times10^{-11}$ & {\boldmath$9.580\times10^{-6}$} & $1.444\times10^{-10}$ \\
\quad fitted order & \multicolumn{2}{c}{$2.35$} & \multicolumn{2}{c}{$2.19$} \\
\quad max residual & \multicolumn{2}{c}{$0.356$} & \multicolumn{2}{c}{$0.374$} \\
\midrule
\multicolumn{5}{l}{\emph{isoAlpha}} \\
32 & $7.748\times10^{-2}$ & $0$ & {\boldmath$8.105\times10^{-2}$} & $0$ \\
64 & $1.615\times10^{-2}$ & $0$ & {\boldmath$1.989\times10^{-2}$} & $0$ \\
128 & $2.052\times10^{-3}$ & $0$ & {\boldmath$2.477\times10^{-3}$} & $0$ \\
256 & $4.956\times10^{-4}$ & $1.465\times10^{-89}$ & {\boldmath$1.261\times10^{-3}$} & $1.153\times10^{-9}$ \\
512 & $9.937\times10^{-5}$ & $2.034\times10^{-10}$ & {\boldmath$1.471\times10^{-4}$} & $1.880\times10^{-10}$ \\
1024 & $3.194\times10^{-5}$ & $2.097\times10^{-9}$ & {\boldmath$4.786\times10^{-5}$} & $1.085\times10^{-9}$ \\
2048 & $1.136\times10^{-5}$ & $9.142\times10^{-10}$ & {\boldmath$1.744\times10^{-5}$} & $1.781\times10^{-8}$ \\
\quad fitted order & \multicolumn{2}{c}{$2.16$} & \multicolumn{2}{c}{$2.07$} \\
\quad max residual & \multicolumn{2}{c}{$0.459$} & \multicolumn{2}{c}{$0.460$} \\
\bottomrule
\end{tabular}
\end{table}
 
The framed error exceeds the unframed error at all seven resolutions.
The ratios range from $1.12$ to $2.82$ for plicRDF and from $1.05$ to $2.54$
for isoAlpha. Across those grids, the fitted refinement exponent decreases from $2.35$ to $2.19$ for plicRDF and from $2.16$ to $2.07$ for isoAlpha.

The maximum absolute log residuals range from $0.356$ to $0.460$. The four
series depart more strongly from a power law than the shear-flow series
in Section~\ref{sec:3d}. The reported exponents summarise the grids tested.
The error increase is consistent with cancellation in the retracing runs.
Rotation also changes mesh alignment, local fluxes and temporal
discretisation error. The paired results do not separate these effects.

The largest recorded boundedness error is $1.8\times10^{-7}$. We monitor
boundedness alongside phase-volume change (Section~\ref{sec:verification})
and fit refinement exponents only to the shape error.

\subsection{A common time grid for paired runs}
\label{sec:protocol}

We use the fixed step from \eqref{eq:dt} for both members of each pair.
Independent adaptive controllers can produce different grids in time.
Near $T/2$ the base amplitude approaches zero while the rigid-frame velocity
persists. In the adaptive vortex comparison at $N=128$, the unframed run
crossed $t\in[3.5,4.5]$ in $35$ steps of up to $5.9\times10^{-2}$. The framed
run took $118$ steps of at most $1.2\times10^{-2}$ over the same interval.

A mesh-independent step cap does not by itself enforce a Courant limit.
Under the inherited adaptive protocol, the observed maximum interface
Courant number in the unframed vortex was $0.23$, $0.32$, $0.52$, $0.65$
and $1.18$ at $N=32$, $128$, $512$, $1024$ and $2048$. The framed runs stayed
at about $0.20$. When a Courant number exceeds one, the swept region can
extend beyond the neighbouring cell. The common grid fixes step counts
and sampling times, while local Courant histories and temporal
discretisation errors can vary with the velocity field.

\section{Three-dimensional shear flow}
\label{sec:3d}

\subsection{Case and vector potential}

We use the shear-flow test of Liovic et al.~\cite{liovic2006} distributed
with the same library \cite{scheufler2019}. A sphere of radius $0.15$ starts
at $(0.5,0.75,0.25)$ in the unit cube, with $N$ cells of equal side length
in each direction. The distributed case uses a box of height $2$. We retain
its initial centre and velocity on the shorter domain. The sampled
trajectories reach $z\simeq0.885$, an estimated distance of $0.115$ below
the upper boundary \cite{bothe2026figshare},
\path{deformation-sphere3d/results/support_extent.txt}. The base field is
\begin{equation}
  \bu_0 =
  \begin{pmatrix}
     \sin(2\pi y)\sin^2(\pi x)\\[2pt]
    -\sin(2\pi x)\sin^2(\pi y)\\[2pt]
    (1-2\rho)^2
  \end{pmatrix},
  \qquad \rho = \lVert (x,y) - \bs\rVert,
  \label{eq:u03d}
\end{equation}
Here $\bs=(0.5,0.5)$ locates the swirl axis horizontally and the amplitude
is $\gamma(t)=\cos(\pi t/3)$ on $[0,3]$. With $a=T/2$, the relation
$\gamma(a+\tau)=-\gamma(a-\tau)$ again gives exact retracing in the base flow.

The field \eqref{eq:u03d} is the sum of a solenoidal horizontal field and
an axial field. The potential $\psi_0\be_z$ generates the horizontal
components, with $\psi_0$ given by \eqref{eq:psi0}. Extend $\psi_0$ by zero
outside the horizontal square (Section~\ref{sec:domain}) and use the radial
formula for the axial component. To obtain its potential, solve
$\tfrac1\rho \,\partial_\rho(\rho A_\theta) = (1-2\rho)^2$ with the regular
choice $A_\theta = \rho/2 - 4\rho^2/3 + \rho^3$. The combined potential is
\begin{equation}
  \bA_0 = \psi_0(x,y)\,\be_z
      + \Bigl(\tfrac12 - \tfrac{4\rho}{3} + \rho^2\Bigr)
        \bigl(\be_z \times (\mathbf{p}_h - \bs)\bigr),
  \label{eq:A03d}
\end{equation}
Here $\mathbf{p}_h$ is the horizontal position. The radial contribution to the
potential is continuously differentiable at the axis, and
$(\nabla\times \bA_0)_z = 1 - 4\rho + 4\rho^2 = (1-2\rho)^2$.
We use the three-dimensional formulas in Section~\ref{sec:streamfun} to
obtain the laboratory potential and the oriented circulation of
Section~\ref{sec:agnostic} to assemble the fluxes.

\subsection{Flux assembly and reference comparison}
\label{sec:projection}

The reference implementation evaluates \eqref{eq:u03d} at face centres
and sets $F_f = \bu_0(\bx_f)\cdot \bS_f$, where $\bx_f$ is the centroid and
$\bS_f$ is the area vector of face $S_f$. The quadrature produces a discrete
divergence residual. The pressure projection \texttt{CorrectPhi} solves a
Poisson equation for a correction potential to enforce
\eqref{eq:discretediv}. With the moving frame, the spatial distribution of
face fluxes changes from step to step. The projection would then require
repeated solves.

Circulation gives discretely solenoidal fluxes directly. Table~\ref{tab:3dproj}
records a separate unframed comparison with the reference implementation.
The terminal errors are close. Isolating the effect of flux assembly in this
auxiliary comparison requires the full run settings. The paired
convergence runs use circulation for both cases and the common time grid
described in Section~\ref{sec:setup}.

\begin{table}[htbp]
\centering
\caption{Separate unframed shear-flow comparison of face-centre quadrature
with Poisson projection and discrete circulation of the analytic vector
potential \eqref{eq:A03d}. The entries give $E_{\text{shape}}(T)$ for
plicRDF.}
\label{tab:3dproj}
\begin{tabular}{rccc}
\toprule
$N$ & quadrature $+$ \texttt{CorrectPhi} & vector potential & difference \\
\midrule
 32 & $4.483\times10^{-3}$ & $4.477\times10^{-3}$ & $0.13\%$ \\
 64 & $1.273\times10^{-3}$ & $1.272\times10^{-3}$ & $0.08\%$ \\
128 & $3.182\times10^{-4}$ & $3.182\times10^{-4}$ & $<0.01\%$ \\
\bottomrule
\end{tabular}
\end{table}

The archived shear-flow convergence runs give the following maxima:
\begin{equation*}
  \max_k \Bigl\lvert \textstyle\sum_{f\in F_k} F_f \Bigr\rvert
  \big/ |\Omega_k| \;\leq\; 2.2\times10^{-11},
  \qquad
  E_{\text{mass}} \;\leq\; 7.7\times10^{-15}
\end{equation*}
\cite{bothe2026figshare}, \path{deformation-sphere3d/results/courant_check.txt}.
The discrete divergence residual grows as the grid is refined. The
calculation divides the flux imbalance by cell volume, which decreases
with refinement, and can therefore amplify round-off. The total
phase-volume change remains below $10^{-14}$.

\subsection{Frame choice and confinement}

The frame rotates about the axis parallel to $\be_z$ through $(0.5,0.5)$.
The horizontal motion has a scalar amplitude, so
Proposition~\ref{prop:band} confines the horizontal projection of the phase
to the streamline band of $\psi_0$ through the sphere's footprint.
This footprint is the disc of radius $0.15$ about $(0.5,0.75)$ used in two
dimensions. The bound in Table~\ref{tab:domain} thus gives
$R_*<0.442$ and places the swept support horizontally inside the unit square.
The axis lies outside the band because its stream-function value
$\psi_0(\bs)$ is greater than the upper band level. Rotation about $\be_z$
preserves the axial coordinate.

The nonnegative flow time for $\bu_0$ is
$(3/\pi)\sin(\pi t/3)$. The centred band gives $\rho<0.442$.
Concavity of $g$ from Section~\ref{sec:domain}, together with $g(0)=0$, gives
$g(a_1^2)+g(a_2^2)\ge g(\rho^2)$ for $\rho<1/2$. By streamline invariance
this yields $\cos^2(\pi\rho(t))\le\pi\psi_0(x_0,y_0)
\le\cos^2(\pi(y_0-0.5))$ for initial horizontal label $(x_0,y_0)$.
It follows that $\rho(t)\ge y_0-0.5$. When $y_0\ge0.625$, the axial
speed is at most $0.75^2$, giving $z\le0.4+(3/\pi)0.75^2<0.938$.
When $0.6\le y_0<0.625$, the initial sphere gives
$z_0\le0.25+\sqrt{0.15^2-0.125^2}<0.333$. The axial speed in this range is
at most $0.8^2$, giving $z<0.945$. The axial speed is nonnegative in
autonomous flow time, so $z\ge0.1$. The phase therefore lies inside the
unit cube throughout the cycle.

\subsection{Interface evolution and convergence}

\begin{figure}[htbp]
\centering
\includegraphics[width=\textwidth]{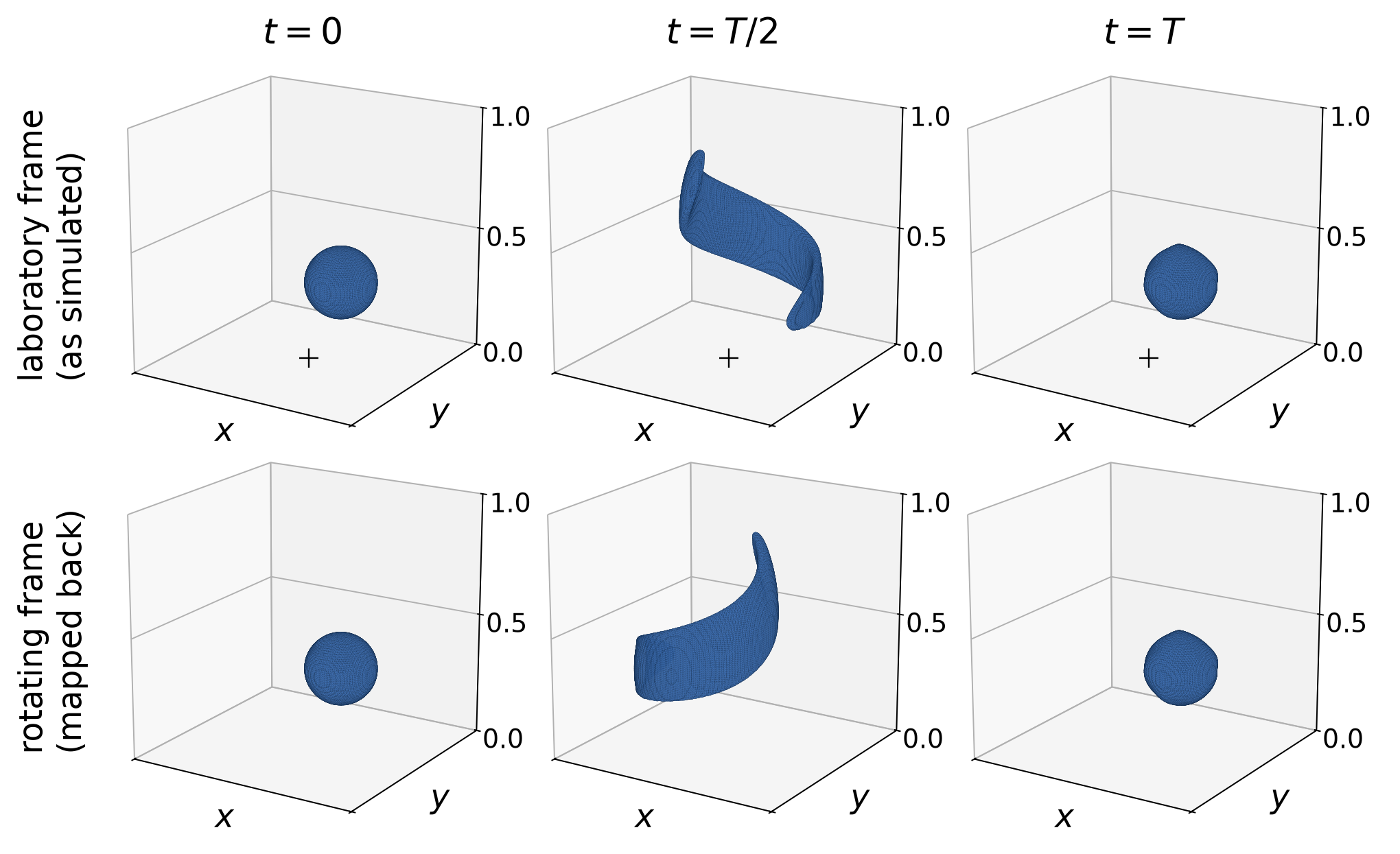}
\caption{Framed three-dimensional shear flow, $\alpha=0.5$ isosurface,
plicRDF, $N=128$, $T=3$, at $t=0$, $T/2$ and $T$. The frame makes one revolution about
the axis parallel to $\be_z$ through $(0.5,0.5)$, marked $+$. The top row
shows the computed interface in laboratory coordinates. The bottom row maps
the same outputs through $E_t^{-1}$ for comparison with
Figure~\ref{fig:static3d}. Data in
\cite{bothe2026figshare}, \protect\path{deformation-sphere3d/results/interface3d_moving_frame.png}.}
\label{fig:moving3d}
\end{figure}

\begin{figure}[htbp]
\centering
\includegraphics[width=\textwidth]{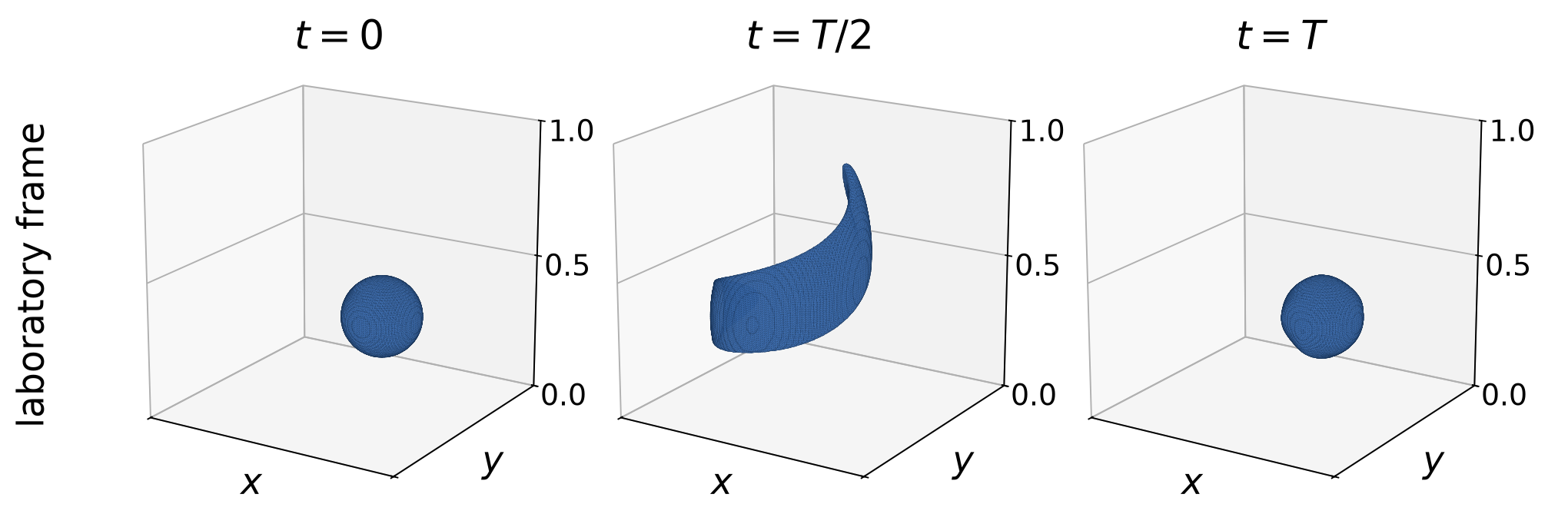}
\caption{Unframed three-dimensional shear flow, $\alpha=0.5$ isosurface,
plicRDF, $N=128$, $T=3$, at $t=0$, $T/2$ and $T$. Data: \cite{bothe2026figshare},
\protect\path{deformation-sphere3d/results/interface3d_non_moving_frame.png}.}
\label{fig:static3d}
\end{figure}

\begin{figure}[htbp]
\centering
\includegraphics[width=\textwidth]{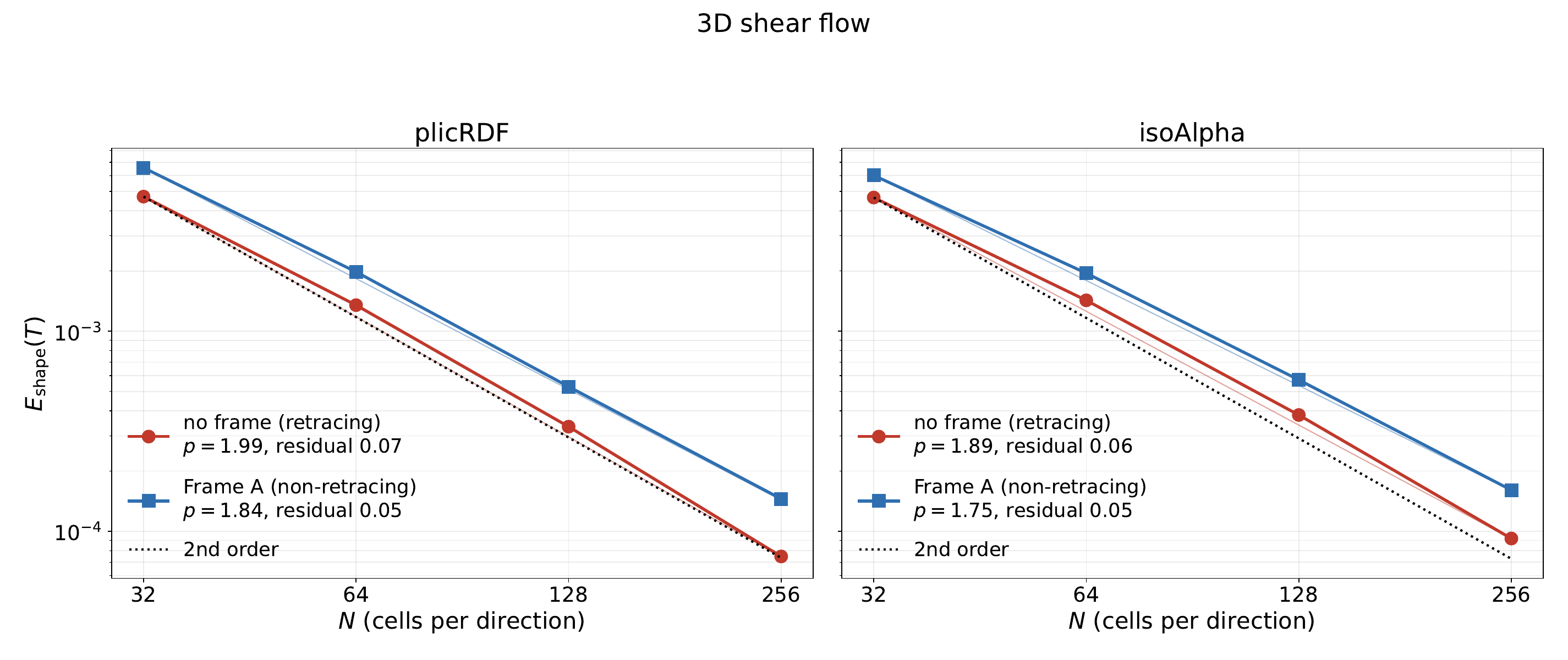}
\caption{Convergence of $E_{\text{shape}}(T)$ for the three-dimensional
shear flow, with and without the superposed frame. The lines show least-squares fits on logarithmic axes. The legend gives
the fitted exponents and maximum absolute log residuals. Data: \cite{bothe2026figshare}, \protect\path{deformation-sphere3d/results/convergence_shape_error.png}.}
\label{fig:conv3d}
\end{figure}

\begin{table}[htbp]
\centering
\small
\setlength{\tabcolsep}{4pt}
\caption{Convergence on the 3D shear-flow test, both reconstruction schemes. The order is the negative least-squares slope of $\ln E_{\text{shape}}$ against $\ln N$ over the whole sequence; the residual is the maximum absolute residual in $\ln E_{\text{shape}}$. Bold: the shape error with the frame exceeds the one without it at the same $N$. Data: \cite{bothe2026figshare}, \protect\path{deformation-sphere3d/results/convergence.csv}.}
\label{tab:conv3d}
\begin{tabular}{r cc cc}
\toprule
& \multicolumn{2}{c}{no frame (retracing)} & \multicolumn{2}{c}{Frame A (non-retracing)} \\
\cmidrule(lr){2-3}\cmidrule(lr){4-5}
$N$ & $E_{\text{shape}}$ & $E_{\text{bound}}$ & $E_{\text{shape}}$ & $E_{\text{bound}}$ \\
\midrule
\multicolumn{5}{l}{\emph{plicRDF}} \\
32 & $4.709\times10^{-3}$ & $8.132\times10^{-20}$ & {\boldmath$6.563\times10^{-3}$} & $2.910\times10^{-14}$ \\
64 & $1.350\times10^{-3}$ & $2.118\times10^{-22}$ & {\boldmath$1.980\times10^{-3}$} & $1.870\times10^{-20}$ \\
128 & $3.333\times10^{-4}$ & $2.220\times10^{-16}$ & {\boldmath$5.283\times10^{-4}$} & $2.220\times10^{-16}$ \\
256 & $7.483\times10^{-5}$ & $2.220\times10^{-16}$ & {\boldmath$1.452\times10^{-4}$} & $5.164\times10^{-14}$ \\
\quad fitted order & \multicolumn{2}{c}{$1.99$} & \multicolumn{2}{c}{$1.84$} \\
\quad max residual & \multicolumn{2}{c}{$0.069$} & \multicolumn{2}{c}{$0.046$} \\
\midrule
\multicolumn{5}{l}{\emph{isoAlpha}} \\
32 & $4.660\times10^{-3}$ & $1.059\times10^{-22}$ & {\boldmath$6.012\times10^{-3}$} & $0$ \\
64 & $1.427\times10^{-3}$ & $2.220\times10^{-16}$ & {\boldmath$1.951\times10^{-3}$} & $2.220\times10^{-16}$ \\
128 & $3.812\times10^{-4}$ & $2.220\times10^{-16}$ & {\boldmath$5.711\times10^{-4}$} & $2.220\times10^{-16}$ \\
256 & $9.197\times10^{-5}$ & $9.733\times10^{-11}$ & {\boldmath$1.606\times10^{-4}$} & $5.789\times10^{-14}$ \\
\quad fitted order & \multicolumn{2}{c}{$1.89$} & \multicolumn{2}{c}{$1.75$} \\
\quad max residual & \multicolumn{2}{c}{$0.065$} & \multicolumn{2}{c}{$0.045$} \\
\bottomrule
\end{tabular}
\end{table}
 
For both reconstructions the framed error exceeds the unframed error at
every resolution. The ratios increase with refinement
(Figure~\ref{fig:conv3d}).

\begin{center}
\begin{tabular}{lcccc}
\toprule
ratio $E_{\text{shape}}^{\text{Frame A}}/E_{\text{shape}}^{\text{no frame}}$
 & $N=32$ & $N=64$ & $N=128$ & $N=256$ \\
\midrule
plicRDF  & 1.39 & 1.47 & 1.59 & 1.94 \\
isoAlpha & 1.29 & 1.37 & 1.50 & 1.75 \\
\bottomrule
\end{tabular}
\end{center}

The fitted exponents are $1.84$ with the frame and $1.99$ without it for
plicRDF, and $1.75$ and $1.89$ for isoAlpha. Over the common grid sequence,
the slope of the fitted log-ratio versus log resolution is the unframed
exponent minus the framed exponent.

The maximum absolute log residual is below $0.070$ for each fit, the
smallest residual range among the three benchmarks. The monotone increase
in both error ratios is specific to this benchmark. It shows how changing
the return path can affect the observed refinement trend when sampling
times and continuum deformation up to rigid motion are matched.

\section{A fully three-dimensional deformation}
\label{sec:leveque}

The shear flow of Section~\ref{sec:3d} consists of a planar vortex and an
axial drift. In the third benchmark we use the LeVeque field
\cite{leveque1996} in the unit-cube geometry of Enright
et al.~\cite{enright2002}. Each velocity component depends on all three
coordinates,
\begin{equation}
  \bu_0 =
  \begin{pmatrix}
     2\sin^2(\pi x)\sin(2\pi y)\sin(2\pi z)\\[2pt]
    -\sin(2\pi x)\sin^2(\pi y)\sin(2\pi z)\\[2pt]
    -\sin(2\pi x)\sin(2\pi y)\sin^2(\pi z)
  \end{pmatrix},
  \label{eq:lev}
\end{equation}
with the same amplitude $\gamma(t)=\cos(\pi t/3)$ on $[0,3]$ and a sphere of
radius $0.15$ at $(0.35,0.35,0.35)$.

\subsection{Vector potential}

To construct a potential $\bA_0 = (A_1, A_2, A_3)$ for \eqref{eq:lev}, set
$A_1 = 0$ and integrate $\partial_x A_3 = -u_{0,2}$ and
$\partial_x A_2 = u_{0,3}$, where $u_{0,i}$ are the components of $\bu_0$.
Choose the integration functions to recover the first component of the
curl. With $S(a)=\sin(2\pi a)$, $C(a)=\cos(2\pi a)$, $s(a)=\sin^2(\pi a)$,
the resulting potential is
\begin{equation}
  \bA_0 = \Bigl(\,0,\;
    \tfrac{1}{2\pi}S(y)\bigl[C(x)s(z) + C(z)\bigr],\;
   -\tfrac{1}{2\pi}C(x)s(y)S(z)\Bigr).
  \label{eq:Alev}
\end{equation}
Differentiation gives $\nabla\times\bA_0=\bu_0$. A numerical check on a
$64^3$ sample grid gives a maximum discrepancy of $6.8\times10^{-16}$.
The check approximates derivatives by the complex step
$\operatorname{Im} f(x+\mathrm{i}h)/h$. For analytic functions real on the
real axis the truncation error is $O(h^2)$. This formula avoids subtractive
cancellation, so a sufficiently small $h$ makes the truncation error
negligible at working precision. The sampled maximum discrepancy for the
shear field is $3.5\times10^{-16}$ \cite{bothe2026figshare},
\path{leveque3d/results/potential_check.txt} and
\path{deformation-sphere3d/results/potential_check.txt}.
The shear samples lie away from the swirl axis, where the radial expression
has a local analytic continuation. The potential's regularity and curl
on the axis follow from the construction in Section~\ref{sec:3d},
independently of this sampled check.

\begin{remark}[Zero extension of the chosen potential]
The vortex potential $\psi_0$ vanishes on the boundary. The chosen potential
\eqref{eq:Alev} has a nonzero tangential component on cube faces. On $x=0$
its second component satisfies
$\tfrac{1}{2\pi}S(y)\left[s(z)+C(z)\right] \neq 0$. The zero extension used in Section~\ref{sec:domain} would therefore
introduce a tangential jump and a surface term in the distribution
$\nabla\times \bA_0$. Formula \eqref{eq:Alev} instead defines a smooth periodic
extension with a solenoidal curl. A zero extension requires a separate
check of the tangential trace.
\end{remark}

A separate unframed comparison on cubic cells with plicRDF gave shape
errors for circulation and projected face-centre fluxes of
$7.720\times10^{-3}$ and $7.743\times10^{-3}$ at $N=32$, and
$3.001\times10^{-3}$ and $3.079\times10^{-3}$ at $N=64$, respectively.
The full run settings for these reference checks are needed to isolate
the effect of the two assemblies. In the paired convergence runs the
recorded maximum divergence residual satisfies
$\max_k\lvert\sum_{f\in F_k} F_f\rvert/|\Omega_k| \leq 2.7\times10^{-11}$ \cite{bothe2026figshare},
\path{leveque3d/results/courant_check.txt}.

\subsection{Confinement}

The base velocity vanishes on the boundary of the unit cube, and
uniqueness keeps the interior base trajectories inside. For the rotated
flow we bound the horizontal support using $f(q)=\sin^2(\pi q)$ and
$I=f(x)f(y)f(z)$. Along each interior base trajectory, the three log-factor
derivatives equal $\pi\gamma(t)\sin(2\pi x)\sin(2\pi y)\sin(2\pi z)$ times
$2,-1,-1$, respectively. Their sum vanishes, so $I$ is invariant.
The initial coordinates lie in $[0.2,0.5]$, which gives $I\ge \sin^6(\pi/5)$.
Since $f(z)\le1$, the same lower bound holds for $f(x)f(y)$. Applying
\eqref{eq:horizontalbound} to the latter product gives
\[
  \rho\le \frac{\sqrt{2}}{\pi}
             \arccos\bigl(\sin^{3/2}(\pi/5)\bigr)
        =0.496666\ldots <0.5 .
\]
The frame preserves $\rho$ and $z$. The horizontal radius bound and
the base confinement in the vertical direction therefore keep the
rotated phase inside the cube. Every point on the axis is a base
equilibrium, and the initial phase is disjoint from it. Uniqueness keeps
each phase label off the axis, as required by the winding argument of
Section~\ref{sec:idea}.

The archived trajectory samples give a sharper estimate, with maximum
horizontal radius near $0.388$ and clearance near $0.112$
\cite{bothe2026figshare}, \path{leveque3d/results/support_extent.txt}. The
largest recorded phase-volume change is $6.5\times10^{-15}$.

\subsection{Interface evolution and grid refinement}

\begin{figure}[htbp]
\centering
\includegraphics[width=\textwidth]{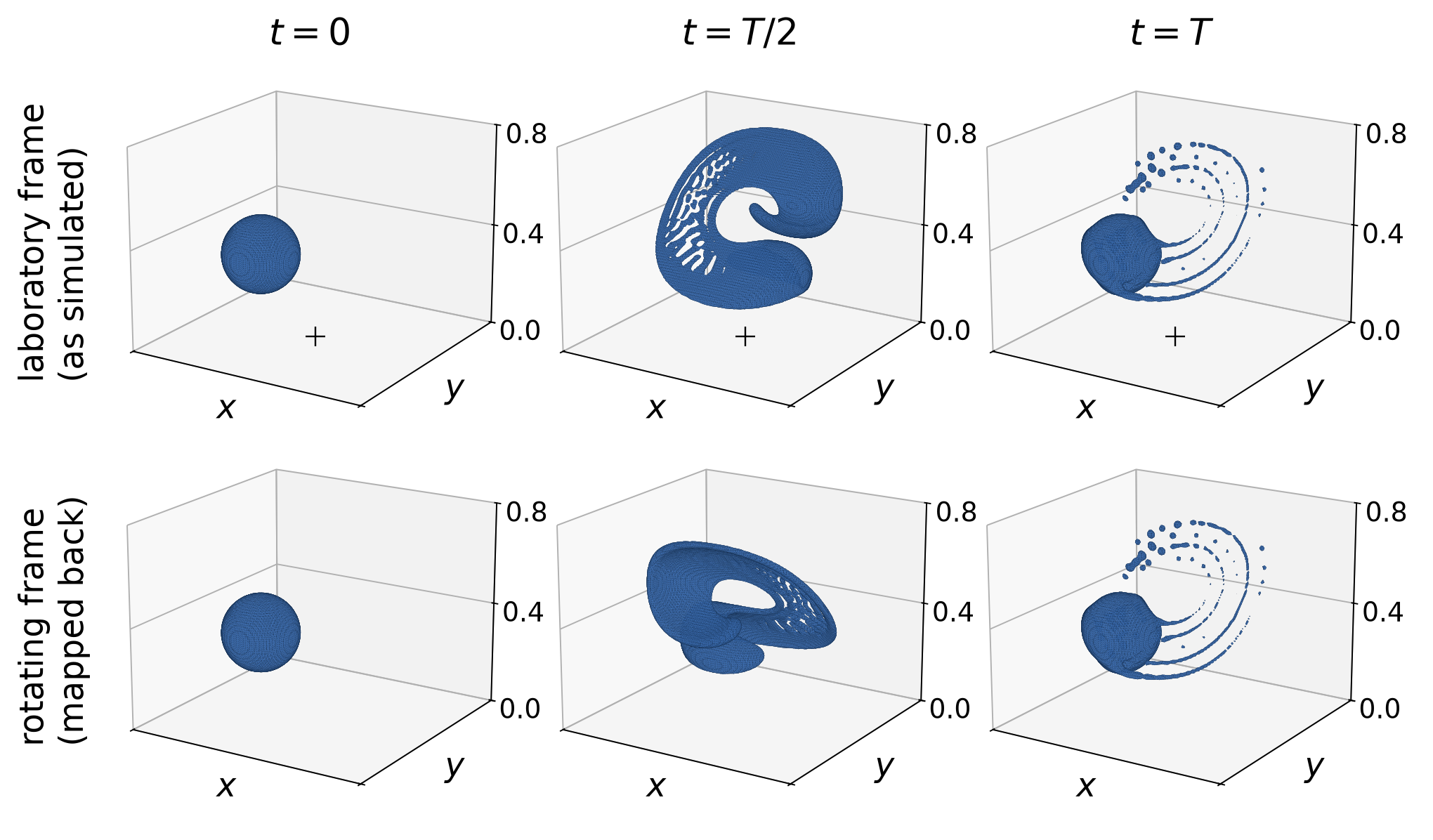}
\caption{Framed LeVeque case, $\alpha=0.5$ isosurface, plicRDF, $N=128$,
$T=3$, at $t=0$, $T/2$ and $T$, with a display cut at $z=0.8$.
The frame rotates once about the axis parallel to $\be_z$ through
$(0.5,0.5)$, marked $+$. The top row shows the interface in laboratory coordinates. The bottom
row maps the same outputs through $E_t^{-1}$. Detached fragments are
visible at $T/2$ and $T$. Data in
\cite{bothe2026figshare}, \protect\path{leveque3d/results/interface3d_moving_frame.png}.}
\label{fig:movinglev}
\end{figure}

\begin{figure}[htbp]
\centering
\includegraphics[width=\textwidth]{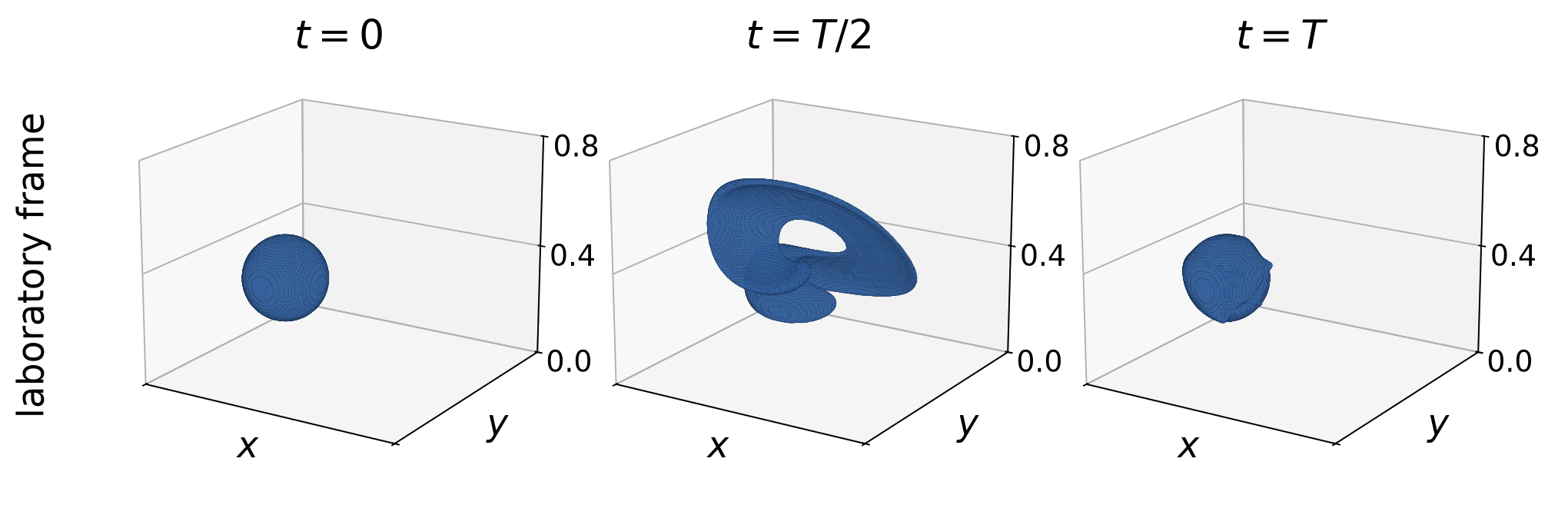}
\caption{Unframed LeVeque case, $\alpha=0.5$ isosurface, plicRDF, $N=128$,
$T=3$, at $t=0$, $T/2$ and $T$; cut at $z=0.8$. Data: \cite{bothe2026figshare},
\protect\path{leveque3d/results/interface3d_non_moving_frame.png}.}
\label{fig:staticlev}
\end{figure}

\begin{figure}[htbp]
\centering
\includegraphics[width=\textwidth]{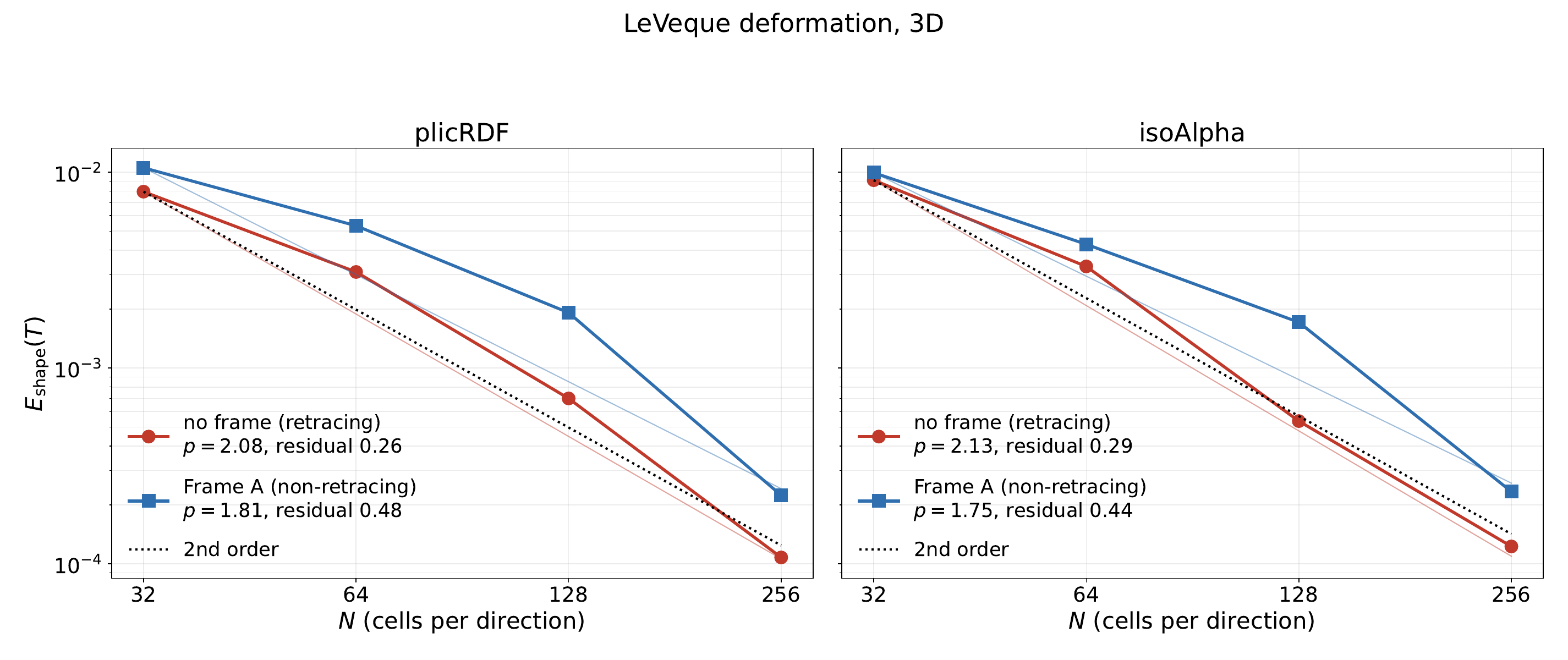}
\caption{Convergence of $E_{\text{shape}}(T)$ for the LeVeque field, with and
without the superposed frame. The lines show least-squares fits on logarithmic axes. The legend gives
the fitted exponents and maximum absolute log residuals.
Data: \cite{bothe2026figshare}, \protect\path{leveque3d/results/convergence_shape_error.png}.}
\label{fig:convlev}
\end{figure}

\begin{table}[htbp]
\centering
\small
\setlength{\tabcolsep}{4pt}
\caption{Convergence on the 3D LeVeque deformation test, both reconstruction schemes. The order is the negative least-squares slope of $\ln E_{\text{shape}}$ against $\ln N$ over the whole sequence; the residual is the maximum absolute residual in $\ln E_{\text{shape}}$. Bold: the shape error with the frame exceeds the one without it at the same $N$. Data: \cite{bothe2026figshare}, \protect\path{leveque3d/results/convergence.csv}.}
\label{tab:convlev}
\begin{tabular}{r cc cc}
\toprule
& \multicolumn{2}{c}{no frame (retracing)} & \multicolumn{2}{c}{Frame A (non-retracing)} \\
\cmidrule(lr){2-3}\cmidrule(lr){4-5}
$N$ & $E_{\text{shape}}$ & $E_{\text{bound}}$ & $E_{\text{shape}}$ & $E_{\text{bound}}$ \\
\midrule
\multicolumn{5}{l}{\emph{plicRDF}} \\
32 & $7.952\times10^{-3}$ & $4.458\times10^{-22}$ & {\boldmath$1.053\times10^{-2}$} & $8.132\times10^{-20}$ \\
64 & $3.094\times10^{-3}$ & $2.220\times10^{-16}$ & {\boldmath$5.326\times10^{-3}$} & $2.220\times10^{-16}$ \\
128 & $6.996\times10^{-4}$ & $2.220\times10^{-16}$ & {\boldmath$1.917\times10^{-3}$} & $2.220\times10^{-16}$ \\
256 & $1.079\times10^{-4}$ & $1.797\times10^{-14}$ & {\boldmath$2.239\times10^{-4}$} & $2.298\times10^{-13}$ \\
\quad fitted order & \multicolumn{2}{c}{$2.08$} & \multicolumn{2}{c}{$1.81$} \\
\quad max residual & \multicolumn{2}{c}{$0.255$} & \multicolumn{2}{c}{$0.484$} \\
\midrule
\multicolumn{5}{l}{\emph{isoAlpha}} \\
32 & $9.100\times10^{-3}$ & $2.220\times10^{-16}$ & {\boldmath$9.939\times10^{-3}$} & $5.294\times10^{-23}$ \\
64 & $3.302\times10^{-3}$ & $8.470\times10^{-22}$ & {\boldmath$4.284\times10^{-3}$} & $1.694\times10^{-21}$ \\
128 & $5.361\times10^{-4}$ & $2.220\times10^{-16}$ & {\boldmath$1.713\times10^{-3}$} & $2.220\times10^{-16}$ \\
256 & $1.228\times10^{-4}$ & $3.559\times10^{-13}$ & {\boldmath$2.340\times10^{-4}$} & $2.529\times10^{-13}$ \\
\quad fitted order & \multicolumn{2}{c}{$2.13$} & \multicolumn{2}{c}{$1.75$} \\
\quad max residual & \multicolumn{2}{c}{$0.287$} & \multicolumn{2}{c}{$0.437$} \\
\bottomrule
\end{tabular}
\end{table}
 
For both reconstructions the framed error exceeds the unframed error at
every resolution (Figure~\ref{fig:convlev}).

\begin{center}
\begin{tabular}{lcccc}
\toprule
ratio $E_{\text{shape}}^{\text{Frame A}}/E_{\text{shape}}^{\text{no frame}}$
 & $N=32$ & $N=64$ & $N=128$ & $N=256$ \\
\midrule
plicRDF  & 1.32 & 1.72 & 2.74 & 2.08 \\
isoAlpha & 1.09 & 1.30 & 3.20 & 1.91 \\
\bottomrule
\end{tabular}
\end{center}

The fitted exponents are $1.81$ with the frame and $2.08$ without it for
plicRDF, and $1.75$ and $2.13$ for isoAlpha. Both error ratios reach their
maximum at $N=128$.

Frame~A increases the maximum absolute log residual from $0.255$ to
$0.484$ for plicRDF and from $0.287$ to $0.437$ for isoAlpha. At $N=128$ the
computed isosurface includes disconnected fragments at $T/2$
(Figure~\ref{fig:movinglev}). Differences in the grid's resolution of the thin
sheet and fragments may contribute to the variation. The fitted exponents
summarise a finite range for which the data depart visibly from a power law.

\section{Using the construction in other solvers}
\label{sec:agnostic}

The benchmark specifies an ambient velocity and an exact terminal
interface. The host method determines how the prescribed velocity enters
its advection update.

\subsection{Velocity-based and flux-based methods}

For a method that samples velocities at points, use the laboratory velocity
\eqref{eq:w} directly. A method that also requires locally balanced carrier
fluxes can construct them from oriented potential circulation.

\subsection{Inputs for potential circulation}

\begin{description}
\item[(R1)] A potential for the base field: a stream function $\psi_0$ in two
  dimensions, a vector potential $\bA_0$ in three, with $\bu_0 = \bJ\nabla\psi_0$ or
  $\bu_0 = \nabla\times \bA_0$. Sections~\ref{sec:basevortex}, \ref{sec:3d} and
  \ref{sec:leveque} give these for the three standard benchmark fields; for a
  new field supply a potential compatible with the intended domain.
\item[(R2)] The mesh vertex coordinates, and for each face its vertex loop in
  the orientation that the code uses for the sign of the face normal. Face
  planarity is unnecessary for the algebraic cancellation. A nonplanar
  face needs a consistently oriented shared spanning surface for the
  interpretation as an exact surface flux.
\item[(R3)] Write access to the face fluxes each time step, and to any sampled velocities required by the host
  advection scheme or its time-step control.
\end{description}

\subsection{Algorithm}

For the separable base field $\bvh(t,\bxi)=\gamma(t)\bu_0(\bxi)$,
proceed at each step with $t$ the step mid-point, frame $\bigl(\bz(t), \bQ(t)\bigr)$,
angular velocity $\bomega(t)$ and base amplitude $\gamma(t)$:

\begin{enumerate}
\item \emph{Potential at the vertices.} For every mesh vertex $p$ with position
  $\bx_p$,
  \begin{equation*}
    \br \leftarrow \bx_p - \bz(t), \qquad
    \bxi \leftarrow \bQ(t)^{\!\top} \br,
  \end{equation*}
  \begin{equation*}
    \bA[p] \leftarrow
      \underbrace{\tfrac12\,\dot \bz(t)\times \bx_p}_{\text{translation}}
      \; \underbrace{-\; \tfrac12 \lVert \br_\perp\rVert^2\,\bomega(t)}_{\text{rotation}}
      \; +\; \underbrace{\gamma(t)\, \bQ(t)\,\bA_0(\bxi)}_{\text{base field}} ,
  \end{equation*}
  with $\br_\perp$ the component of $\br$ orthogonal to $\bomega$ and the
  rotation term set to zero when $\bomega=\mathbf{0}$. In the
  planar implementation use $\bA=\Psi\,\be_y$, with $\Psi$ from
  \eqref{eq:Psi}.
\item \emph{Fluxes by circulation.} For every face $S_f$, internal and boundary
  alike,
  \begin{equation*}
    F_f \leftarrow \sum_{(p_1,p_2)\,\in\,\partial S_f}
      \tfrac12\bigl(\bA[p_1] + \bA[p_2]\bigr)\cdot\bigl(\bx_{p_2} - \bx_{p_1}\bigr),
  \end{equation*}
  the sum running over the oriented edges of the face loop.
\item \emph{Cell velocity, if required.} For every cell centre, evaluate
  \eqref{eq:w} directly when the host scheme uses cell-centred velocities.
\end{enumerate}

Step~2 enforces \eqref{eq:discretediv}: the edge contributions cancel for any
$\bA$, so \eqref{eq:discretediv} holds to round-off regardless of the quadrature
error of the edge integrals, provided each cell boundary uses a consistent
oriented incidence and common edge values. Boundary fluxes must retain that
balance.

The algorithm consists of one loop over the vertices and one over the faces,
without a linear solve or a global reduction. A vertex potential depends only on
the vertex position and prescribed frame data. Parallel implementations must
use consistent shared-face fluxes and orientations. The cost is
$O(N_v) + O\!\left(\sum_f n_f\right)$, with $n_f$ the number of vertices of the
face $S_f$. Compared with projecting a face-centre quadrature
(Section~\ref{sec:projection}), the algorithm obtains balanced fluxes directly from the analytic potential
and avoids a linear solve.

\subsection{Orientation and flux balance}

Orient each face loop consistently with the signed face normal used by the
host code. With the outward cell signs, monitor
\begin{equation}
  \max_k\left(
    \Bigl\lvert\textstyle\sum_{f\in F_k}F_f\Bigr\rvert/|\Omega_k|
  \right).
  \label{eq:solencheck}
\end{equation}
Compare this residual with the scale of the face terms after dividing by cell
volume. Reversing all face-loop signs leaves the residual unchanged. A
final-time return can also survive a sign error, as in a full revolution
with the wrong sense of rotation. A separate signed check can compare a
face flux with its analytic value for a uniform translation or check a
material displacement at an intermediate time.

\subsection{Implementation checks}
\label{sec:ladder}

The following checks distinguish flux assembly from errors in
interface transport.
\begin{enumerate}
\item \emph{Carrier balance and orientation.} Monitor
  \eqref{eq:solencheck} and check an oriented reference flux or displacement.
\item \emph{Null frame.} With $\bQ\equiv\bI$ and $\bz\equiv\mathbf{0}$,
  compare against the reference implementation. The two paths should agree
  to round-off when they assemble the same discrete inputs. With different
  flux quadratures, compare their discretization errors as in
  Section~\ref{sec:projection}.
\item \emph{Frame only.} With $\gamma\equiv0$ the field is a rigid motion
  with a known solution. The resulting error gives the scale of the
  rigid-motion control on the chosen mesh.
\item \emph{Inverse mapping.} Map intermediate numerical interfaces back
  with $E_t^{-1}$ and compare them in the base coordinates. The exact
  configurations agree. The numerical difference reflects the errors of
  both runs and any error in their plotted reconstructions.
\end{enumerate}
Use the same time grid in a pair and check the swept phase support and
recorded volume drift.

\subsection{Choosing a frame}

A returning frame can be chosen to control how the phase moves relative to
the mesh. For example, for a pure translation
$E_t(\by)=\by+\varepsilon\bz_*(t)$ with $\bz_*(0)=\bz_*(T)=\mathbf{0}$,
the mirrored-time gap is
$|\varepsilon|\|\bz_*(T-t)-\bz_*(t)\|$ for every label. Small $|\varepsilon|$
gives nearly retraced paths. A full turn or a combination of rotation
and translation can explore different orientations and locations while
keeping the exact return. Check separation on the relevant labels using
\eqref{eq:noretrace}. Also bound or estimate the swept support and verify
that it fits the domain, as in Sections~\ref{sec:domain} and~\ref{sec:leveque}.

\section{Conclusions}

\label{sec:conclusions}

A cyclic change of frame varies the path of a deforming phase across a
stationary mesh. The base flow and the frame each return, so the phase
returns exactly with its deformation history preserved up to rigid
placement. A small returning motion yields almost retraced paths.
For a full rotation about a fixed centre or axis, material labels that
stay off it have nonzero winding, which excludes complete geometric
retracing.
For a divergence-free base flow, potential circulation gives locally
balanced carrier fluxes for geometric VOF.

With the full-turn frame and a shared time grid, the final shape errors
were larger in all 30 pairs, by factors of $1.05$ to $3.20$ and a median
factor of $1.50$. The fitted refinement exponents were smaller with the
frame in all six comparisons.
The shear test showed the most regular convergence over the tested grids
and a monotone increase in the framed-to-unframed error ratio. The vortex
and fully three-dimensional deformation showed more variation across
resolutions. The results establish a dependence on the return path for
these cases. Both reversal cancellation and the changed motion relative
to the grid can influence the differences, and the paired tests do not
isolate their contributions. As the one-dimensional examples show,
breaking mirrored-time retracing alone does not guarantee a larger
terminal error.

\appendix
\section{Exact return of the time-reversed base flow}
\label{app:proof}

Let $\Omega\subseteq\R^d$, $T>0$ and $a=T/2$, and let
$\bu:[0,a]\times\Omega\to\R^d$ be the first-half field of
Section~\ref{sec:step1}. Assume first-half trajectories exist as absolutely
continuous curves in $\Omega$ over $[0,a]$. On $\R^d$, a sufficient setting
for existence and uniqueness is measurability in time, an integrable bound
at one spatial reference point, and the spatial Lipschitz bound
\eqref{eq:lip} with nonnegative $L\in L^1(0,a)$. On a restricted domain
the existence interval must also cover the relevant trajectories.
The reflected first-half curve solves the second-half equation and stays
in the same base domain. The uniqueness estimate below then gives the
terminal identity on the admissible initial positions.

The base field $\bvh$ is defined by \eqref{eq:vdef}. Its value at the single
instant $t=a$ does not affect the integral form of the differential equation,
so $\bvh$ may jump there; all derivatives of trajectories below hold almost
everywhere, or one-sided where appropriate. The flow map $\Phih_s^t$ of $\bvh$
satisfies, for starting points and times for which the map is defined,
\begin{equation}
  \frac{\dd}{\dd t}\Phih_s^t(\by_s) = \bvh\bigl(t,\Phih_s^t(\by_s)\bigr),
  \qquad \Phih_s^s(\by_s) = \by_s .
  \label{eq:app:flow}
\end{equation}
Fix an arbitrary $\by_0\in\Omega$ and write
\begin{equation}
  \by(t) = \Phih_0^t(\by_0), \qquad
  \frac{\dd\by}{\dd t} = \bvh\bigl(t,\by(t)\bigr), \qquad \by(0) = \by_0 .
  \label{eq:app:traj}
\end{equation}

\subsection{The two mirrored functions satisfy the same equation}

For $0<\tau<a$, the chain rule applied to the position at time $a-\tau$ gives
\begin{equation}
  \begin{split}
  \frac{\dd}{\dd\tau}\by(a-\tau)
  &= \left.\frac{\dd\by}{\dd t}\right|_{t=a-\tau}\frac{\dd(a-\tau)}{\dd\tau}
   = -\bvh\bigl(a-\tau,\by(a-\tau)\bigr)\\
  &= -\bu\bigl(a-\tau,\by(a-\tau)\bigr),
  \end{split}
  \label{eq:app:minus}
\end{equation}
where the first branch of \eqref{eq:vdef} applies because $a-\tau<a$. For the
position at time $a+\tau$, the second branch applies, and $T-a=a$ gives
\begin{equation}
  \begin{split}
  \frac{\dd}{\dd\tau}\by(a+\tau)
  &= \left.\frac{\dd\by}{\dd t}\right|_{t=a+\tau}\frac{\dd(a+\tau)}{\dd\tau}
   = \bvh\bigl(a+\tau,\by(a+\tau)\bigr)\\
  &= -\bu\bigl(T-a-\tau,\by(a+\tau)\bigr)
   = -\bu\bigl(a-\tau,\by(a+\tau)\bigr).
  \end{split}
  \label{eq:app:plus}
\end{equation}
Equations \eqref{eq:app:minus} and \eqref{eq:app:plus} show that
$\by(a-\tau)$ and $\by(a+\tau)$, both viewed as functions of $\tau$, satisfy
the same differential equation \eqref{eq:sameode}. At $\tau = 0$ both are the
midpoint position,
\begin{equation}
  \left.\by(a-\tau)\right|_{\tau=0} = \by(a) = \left.\by(a+\tau)\right|_{\tau=0} .
  \label{eq:app:init}
\end{equation}
The two functions therefore solve the same initial-value problem on $[0,a]$. To
conclude that they coincide, it remains to show that this problem has a unique
solution.

\subsection{Lipschitz continuity and Gr\"{o}nwall's inequality imply uniqueness}

Let $\by_1(\tau)$ and $\by_2(\tau)$ be two solutions of \eqref{eq:sameode} with
$\by_1(0) = \by_2(0)$,
\begin{equation}
  \frac{\dd\by_1}{\dd\tau} = -\bu(a-\tau,\by_1), \qquad
  \frac{\dd\by_2}{\dd\tau} = -\bu(a-\tau,\by_2),
  \label{eq:app:two}
\end{equation}
and let $\boldsymbol\delta(\tau) = \by_1(\tau) - \by_2(\tau)$ be their
difference. Differentiating the definition and inserting \eqref{eq:app:two}
gives
\begin{equation}
  \frac{\dd\boldsymbol\delta}{\dd\tau}
  = \frac{\dd\by_1}{\dd\tau} - \frac{\dd\by_2}{\dd\tau}
  = -\bu(a-\tau,\by_1) + \bu(a-\tau,\by_2)
  = -\bigl[\bu(a-\tau,\by_1) - \bu(a-\tau,\by_2)\bigr].
  \label{eq:app:ddelta}
\end{equation}
The squared Euclidean norm is $\lVert\boldsymbol\delta\rVert^2 =
\boldsymbol\delta\cdot\boldsymbol\delta$, and the product rule gives
\begin{equation}
  \frac{\dd}{\dd\tau}\lVert\boldsymbol\delta\rVert^2
  = 2\,\boldsymbol\delta\cdot\frac{\dd\boldsymbol\delta}{\dd\tau}
  = -2\,\boldsymbol\delta\cdot
    \bigl[\bu(a-\tau,\by_1) - \bu(a-\tau,\by_2)\bigr].
  \label{eq:app:dnorm}
\end{equation}
With $-b\leq\lvert b\rvert$ for every real $b$ and the Cauchy--Schwarz
inequality $\lvert\mathbf p\cdot\mathbf q\rvert\leq\lVert\mathbf p\rVert\,
\lVert\mathbf q\rVert$,
\begin{equation}
  \frac{\dd}{\dd\tau}\lVert\boldsymbol\delta\rVert^2
  \leq 2\,\lVert\boldsymbol\delta\rVert\,
    \bigl\lVert\bu(a-\tau,\by_1) - \bu(a-\tau,\by_2)\bigr\rVert .
  \label{eq:app:cs}
\end{equation}
The Lipschitz condition \eqref{eq:lip} at the time $t = a-\tau$ bounds the last
factor by $L(a-\tau)\lVert\by_1-\by_2\rVert = L(a-\tau)\lVert\boldsymbol\delta
\rVert$, so
\begin{equation}
  \frac{\dd}{\dd\tau}\lVert\boldsymbol\delta(\tau)\rVert^2
  \leq 2L(a-\tau)\,\lVert\boldsymbol\delta(\tau)\rVert^2 .
  \label{eq:app:diffineq}
\end{equation}
The differential form of Gr\"{o}nwall's inequality states that if $f\geq0$ is
absolutely continuous and $f'(\tau)\leq g(\tau)f(\tau)$ almost everywhere, with
$g$ integrable, then
\begin{equation}
  f(\tau) \leq f(0)\exp\!\left(\int_0^\tau g(s)\,\dd s\right).
  \label{eq:app:gronwall}
\end{equation}
The function $f = \lVert\boldsymbol\delta\rVert^2$ is absolutely continuous
because $\by_1$ and $\by_2$ are. With $g(\tau) = 2L(a-\tau)$,
\begin{equation}
  \lVert\boldsymbol\delta(\tau)\rVert^2 \leq \lVert\boldsymbol\delta(0)\rVert^2
  \exp\!\left(2\int_0^\tau L(a-s)\,\dd s\right),
  \label{eq:app:bound}
\end{equation}
where the integral is finite because $L\in L^1(0,a)$. The equal initial values
give $\boldsymbol\delta(0) = \mathbf 0$, so \eqref{eq:app:bound} gives
$\lVert\boldsymbol\delta(\tau)\rVert^2 \leq 0$. A squared norm is nonnegative,
hence
\begin{equation}
  \lVert\boldsymbol\delta(\tau)\rVert = 0, \qquad
  \by_1(\tau) = \by_2(\tau) \quad\text{for every } \tau\in[0,a],
  \label{eq:app:unique}
\end{equation}
which is the uniqueness of the solution of \eqref{eq:sameode}.

\subsection{Retracing and the identity of the final flow map}

The functions $\by(a-\tau)$ and $\by(a+\tau)$ solve \eqref{eq:sameode} with the
same initial value \eqref{eq:app:init}. By \eqref{eq:app:unique} they coincide,
which is the retracing \eqref{eq:retrace},
$\by(a+\tau) = \by(a-\tau)$ for $0\leq\tau\leq a$. For $\tau = a$,
\begin{equation}
  \by(a+a) = \by(a-a) \quad\Longleftrightarrow\quad \by(T) = \by(0) = \by_0 .
  \label{eq:app:return}
\end{equation}
Since $\by(t) = \Phih_0^t(\by_0)$ and $\by_0$ was arbitrary,
$\Phih_0^T(\by_0) = \by_0$ for every admissible $\by_0$, that is,
$\Phih_0^T = \mathrm{Id}$.

The retracing \eqref{eq:retrace} is stronger than the return: every material
point traverses its first-half trajectory in reverse order.
Section~\ref{sec:step2} gives the condition under which a returning frame
breaks this symmetry for a given material point.

\section*{CRediT authorship contribution statement}

\textbf{Dieter Bothe:} Conceptualization, Data curation, Formal analysis,
Investigation, Methodology, Resources, Software, Validation, Visualization,
Writing -- original draft, Writing -- review \& editing.
\textbf{Tomislav Mari\'{c}:} Data curation, Formal analysis, Investigation,
Methodology, Software, Validation, Visualization, Writing -- original draft,
Writing -- review \& editing.

\section*{Declaration of competing interest}

The authors declare that they have no known competing financial interests or
personal relationships that could have appeared to influence the work reported
in this paper.

\section*{Data availability}

The code and data supporting this article are openly available in the
archive \cite{bothe2026figshare}. It contains the \texttt{TwoPhaseFlow}
source code at the tag \texttt{arxiv-nonretracing-advection-v1}
\cite{maric2026twophaseflow}, the Snakemake workflows of the three studies,
the error time series and case settings of every run, all generated results,
and the scripts for Figure~\ref{fig:points} and
Section~\ref{sec:cancel1d}. Figure and table captions name the file each
was produced from. The tables and convergence figures can be regenerated
from the archived run data without OpenFOAM
(\path{verification/postprocess.py}). The studies themselves are rerun
with one command each, as described in the archive's \path{README.md}.

\section*{Acknowledgements}

The authors gratefully acknowledge the computing time provided to them on the
high-performance computer Lichtenberg at the NHR Centers NHR4CES at TU
Darmstadt. This is funded by the Federal Ministry of Education and Research,
and the state governments participating on the basis of the resolutions of the
GWK for national high performance computing at universities. Tomislav Mari\'{c}
gratefully acknowledges the support of the Athene Young Investigator Programme
of the Technical University of Darmstadt.

\section*{Declaration of generative AI and AI-assisted technologies in the
research and manuscript preparation process}

Both authors used generative AI tools in this work, and we state here how.
Dieter Bothe used ChatGPT (OpenAI) to help derive the transformation of
Section~\ref{sec:method}, to write a first Python implementation of it and to
draft its first description. He later used it for mathematical review and
editorial revision of the manuscript, including the analytic illustration of
Figure~\ref{fig:points} and its script.
Tomislav Mari\'{c} used Claude Code (Anthropic), an agentic coding assistant
based on Claude models, to write the implementation in \texttt{TwoPhaseFlow}
(the flux assembly from the stream function and the vector potentials, the
moving-frame classes, the Snakemake workflows, the post-processing and the
data archive), to run and monitor the simulations on the Lichtenberg cluster,
to draft and edit the text from the authors' notes and review comments, and
to search and check the cited literature.
The AI tools are not authors. The generated code was checked by the tests
reported in Section~\ref{sec:verification} and by the grid-refinement
studies. The authors checked the derivations and the numbers in the text
against the code and the archived results. They reviewed and edited all
content and take full responsibility for the published article.

\bibliographystyle{abbrvnat}
\bibliography{literature}

\end{document}